# Post-Quantum Cryptography Algorithm's Standardization and Performance Analysis

Dr. Manish Kumar
*Assistant Professor*
*Department of Master of Computer Applications*
*M S Ramaiah Institute of Technology*
*Bangalore – 54, INDIA*
*Email: - manishkumarjsr@yahoo.com*

**Abstract: -** Quantum computer is no longer a hypothetical idea. It is the world's most important technology and there is a race among countries to get supremacy in quantum technology. It's the technology that will reduce the computing time from years to hours or even minutes. The power of quantum computing will be a great support for the scientific community. However, it raises serious threats to cybersecurity. Theoretically, all the cryptography algorithms are vulnerable to attack. The practical quantum computers, when available with millions of qubits capacity, will be able to break nearly all modern public-key cryptographic systems. Before the quantum computers arrive with sufficient 'qubit' capacity, we must be ready with quantum-safe cryptographic algorithms, tools, techniques, and deployment strategies to protect the ICT infrastructure. This paper discusses in detail the global effort for the design, development, and standardization of various quantum-safe cryptography algorithms along with the performance analysis of some of the potential quantum-safe algorithms. Most of the quantum-safe algorithms need more CPU cycles, higher runtime memory, and large key size. The objective of the paper is to analyze the feasibility of the various quantum-safe cryptography algorithms.

**Keywords: -** *Quantum Cryptography, Code Based Cryptography, Multivariate Quadratic Equations Based Cryptography, Hash-Based Cryptography, Isogeny Based Cryptography, Lattice-Based Cryptography*

## 1. Introduction

Cryptography is one of the most important pillars of cybersecurity. Cryptography is an art of rendering technique and science of securing the information, where the information can be interpreted only by the sender and intended recipient. Most of the modern cryptography algorithms are based on complex mathematical functions. These mathematical functions are a kind of one-way function which is easy to solve in one direction but very hard to solve in the reverse direction. The security of these cryptography algorithms is based on the fact that existing computing recourse will take hundreds of years to break these algorithms.

Quantum computers, however, can easily break most of the modern cryptography algorithms in feasible time complexity. Quantum computing is a technology based on the principles of quantum theory which is much faster than classical computing techniques. The traditional cryptography algorithms are completely vulnerable to quantum computers. Transitioning to the quantum computer will make the existing IT infrastructure completely unsafe and it requires the design and development of quantum-safe or quantum-resistant cryptography algorithms. The objective of this paper is to discuss various cryptography techniques and potential quantum-safe cryptography algorithms along with the performance analysis of the algorithms.

The paper is divided into eight sections. Section 2 gives general information about cryptography followed by section 3 introducing quantum computers. Section 4 discusses the post-quantum encryption problems. Many countries and professional organizations are putting effort to standardize the quantum-safe algorithms which are discussed in section 5. The National Institute of Standards and Technology (NIST) has initiated an intensive process to standardize the post-quantum encryption algorithm. Some of the potential quantum-safe algorithms which are announced by the NIST in 3rd round of the selection process have been discussed in section 6 along with the performance analysis of the algorithms based on the Open Quantum Safe (OQS) project implementation. The overall observation, implementation feasibility, and migration challenges are discussed in section 7 followed by a concluding remark in section 8.

## 2. Cryptography –The Building Blocks of Information Security

Cryptography has a very long and interesting history. Cryptographic techniques are known to have existed in ancient times, and most early civilization appears to have used cryptography to some degree. Many documents claim the use of cryptography techniques to protect sensitive information around 3500 years ago when Mesopotamian scribe employed cryptography to conceal a formula for pottery glaze, which was used on clay tablets. Although cryptography has a long and enriched history, it wasn't until World War-I, where cryptography and cryptanalysis (the science of finding weaknesses in cryptosystems) played a major role.

Today, information security without cryptography is unimaginable thought. Cryptography is used in various ways to protect the confidentiality and integrity of the information. Information security hardware and software products implement cryptography in many creative ways with the sole purpose of providing security. The cryptographic algorithms are mathematical functions that transform the user data in an encrypted form using the variable called 'key'. Sharing the 'key' with the recipient through a secured channel and protecting the 'key' is most vital for the continued security of the data [54].

Cryptographic algorithms are broadly classified into two categories i.e., Symmetric Cryptographic algorithms and Asymmetric Cryptographic Algorithms.

i) **Symmetric Cryptography Algorithms: -** In symmetric cryptography algorithms, the sender and receiver use the same key to encrypt and decrypt the data. Symmetric key algorithms are very efficient and easy to implement. However, key management in symmetric cryptography is a challenging issue. The key must be shared among the two parties before initiating the communication. Symmetric ciphers are mainly susceptible to chosen-plaintext attack, known-plaintext attack, linear cryptanalysis, and differential cryptanalysis.

ii) **Asymmetric Cryptography Algorithm: -** In an asymmetric cryptography algorithm, the sender and receiver use a pair of keys. One of the keys is kept secret i.e., called a private key and the other key is made public called as public key. An asymmetric cryptography algorithm is also known as a public-key algorithm. The sender encrypts the data using the public key and the receiver decrypts the data using a private key. Generally, asymmetric cryptography algorithms are resource-intensive and require more processing time and power. Asymmetric cryptography algorithms satisfy the requirements for digital signatures, cryptographic key establishment, and many more security services. The public-key ciphers are mainly vulnerable to chosen-ciphertext attacks, man-in-the-middle attacks, and some specific attacks related to the underlying hard problems.

Symmetric and asymmetric, both the algorithms are vulnerable to brute force attack [23][49]. In a brute force attack, the attacker tries all the possible keys. The resistance of the algorithms depends on the key size and computational power required to generate and apply all the possible keys to break the encryption.

Cryptography does not only secure the data but is also used for confidentiality, authentication, integrity, and accessibility. There are various implementations of cryptography algorithms in our daily life. For example, the use of SSL/TLS to provide secure internet browsing, digital signature for authentication and validation, electronic money transaction, secure data storage, etc. [17].

There is no cryptography algorithm in the world that is 100% secure. The strength of any cryptography algorithm depends on the computational time and effort required to break the algorithm. Cryptanalysis and brute force techniques are commonly used for breaking the cryptography algorithm. The cryptography algorithm assumes to be secure if it is proven mathematically that to break the algorithm, requires huge computational resources and time which is practically not feasible.

There is continuous growth in computational power and accordingly, the cryptography algorithms are also modified to maintain the strength. Most of the cryptography algorithms used today are resilient against a classical computer. A Quantum computer is new computation technology that is completely different than a classical computer. A Quantum computer is much faster than a classical computer.

Quantum computing is the next revolutionary technology. It's the technology that will have the ability to reduce the computing time from years to hours or even minutes. The power of quantum computing will be a great support for the scientific community. However, it raises serious cybersecurity threats. Most of the existing cryptography algorithms are theoretically vulnerable to quantum computing. A practical quantum computer with sufficient numbers of qubits will be able to break all the modern public-key cryptographic systems [12].

## 3. Quantum Computing

A Quantum computer is no longer a hypothetical idea. As quoted by many experts, it is the world's most important technology and there is a race among countries to get supremacy in quantum technology. US, China, France, UK, Germany, and Russia are front runners in the race, whereas other countries are putting their best effort to join the league. Race to hold control over "Quantum Computing" technology is not only limited to country, it is significantly driven by the top technology giant like Microsoft, IBM, Google, D-Wave, Toshiba, etc.

Quantum computing got popular after the publication of the article "Simulating Physics with Computer" by American theoretical physicist Feynman [28]. In the article, Feynman suggested the use of quantum states for calculations.

Quantum computing is an application of a quantum mechanism that uses a quantum phenomenon to perform computation. A Quantum computer is a device that performs quantum computing. It manipulates the states of qubits in a controlled way to perform algorithms.

A qubit (or quantum bit) is the quantum-mechanical analogue of a classical bit. In classical computers information is encoded in a bit, where each bit can be either zero or one. In quantum computing, the information is encoded in qubits. The state of the qubits is written as $|0\rangle$ and $|1\rangle$. The qubits can be simultaneously both at 0 and 1. Quantum mechanism is a strange phenomenon. Quantum computers are built using the following features of quantum states: -

i) **Superposition: -** Quantum systems can exist in two states at once. A qubit can be in 0 and 1 at the same time. When the measurement is performed, the qubit collapses to either zero or one.

ii) **Entanglement: -** It's a quantum mechanical phenomenon where the state of entangled particles can be described with reference to each other. Measurement performed on one entangled particle will immediately influence the other entangled particle irrespective of the distance among the entangled particle.

iii) **Interference:** - The fundamental idea in quantum computing is to control the probability of qubits collapsing into a particular measurement state. Quantum interference, a by-product of superposition, allows controlling the measurement of a qubit toward a desired state or set of states.

### *3.1 The Timescale of Quantum Computing Development*

The quantum computing field is growing rapidly. Globally there is exponential growth in this field. Building a quantum computer with higher qubits and precise error control is the sole objective of the researchers. There are remarkable achievements in the last 20 years in this field (Figure 1). MIT, Oxford,

Berkely, and IBM could develop a 2-qubit quantum computer early in 1998. Google developed a 72-qubit quantum computer in 2018. Rigetti in 2019, announced to develop a 128-qubit quantum computer within a year [31][36].

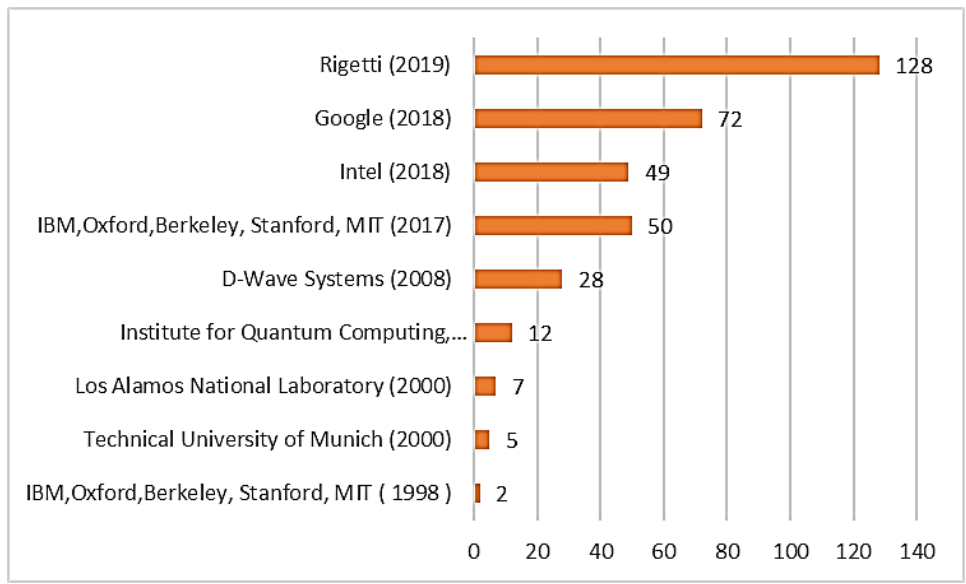

*Figure 1*: - *Quantum Computing Growth in Last 20 Years (Between 1998 to 2019)*

The development of Quantum Computer can be divided into three generation

i) **First Generation: -** The first-generation quantum computers are developed at the early stage was for non-commercial use. These models were built for proof of concept with low to medium complexity.

ii) **Second Generation: -** Many organizations who got the breakthrough in their initial research and possessed the necessary hardware infrastructure could develop the quantum computer with a higher number of qubits and complexity. The second-generation quantum computers are solely designed and developed for commercial applications and high-end research focused on improved scalability and speed. These quantum computers can be rented out for higher computing demand just like cloud computing to serve on a demand basis.

iii) **Third Generation: -** Third generation will be true quantum supremacy as the exponential growth and development will bring down the hardware cost. The quantum computer will be affordable and easily accessible to the mass. The third-generation quantum computer will bring the viable solution across a wide variety of non-commercial applications and it will outperform the classical computer and applications.

There are many technological challenges for the development of Quantum Computer with a higher number of qubits. Initially, the breakthrough research in quantum computing was considerably slow and took significant time to realize the working model. However, in the last few years, there is an unprecedented development. Scientists around the world are addressing various challenging issues to develop an affordable quantum computer with higher accuracy[8][33]. Gartner in 2018 has published the estimated timeline for physical qubits capacity and quantum computer application in real-life (Figure 2). If the Gartner timeline is to be believed, society will witness quantum supremacy very soon. Though envisioning the future shows a very good achievement for the computation field, it's a sign of a storm for cyber security.

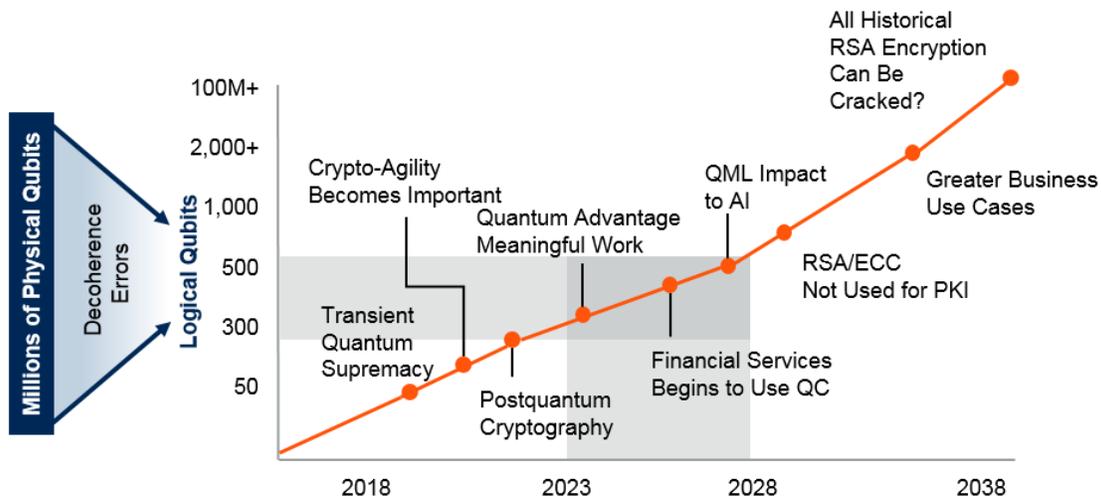

*Figure 2: - Qubit Timeline Estimate (Source Gartner [53])*

The development of quantum technology is good news for humanity as it will uplift our life. However, it also raises a serious threat to cybersecurity, requiring a change in how we encrypt our data. As per MIT Technology Review highlights, 20 million-qubit quantum computers could break a 2048-bit number in a mere 8 hours [5]. Though the 20 million-qubit quantum computers currently do not exist, we need to be prepared and stay ahead of the threat. We can't wait until those powerful quantum computers start breaking our encryption, it will be too late. It's the time to critically analyze the post-quantum threat and prepare accordingly [55][59]. We need to identify the research gap and make a strategic plan to breach the gap.

## 4. Post Quantum Encryption Problem

Encryption algorithms transform the information in such a way that it is difficult to interpret it unless and until it is decrypted using the decryption key [18][56]. Most of the encryption algorithms are based on the complex mathematical function (Figure 3) which is easy to compute in one direction but very difficult to compute in the reverse direction[15][19]. These algorithms are known as one-way functions. RSA is one of the most commonly used encryption algorithms which makes use of the one-way function.

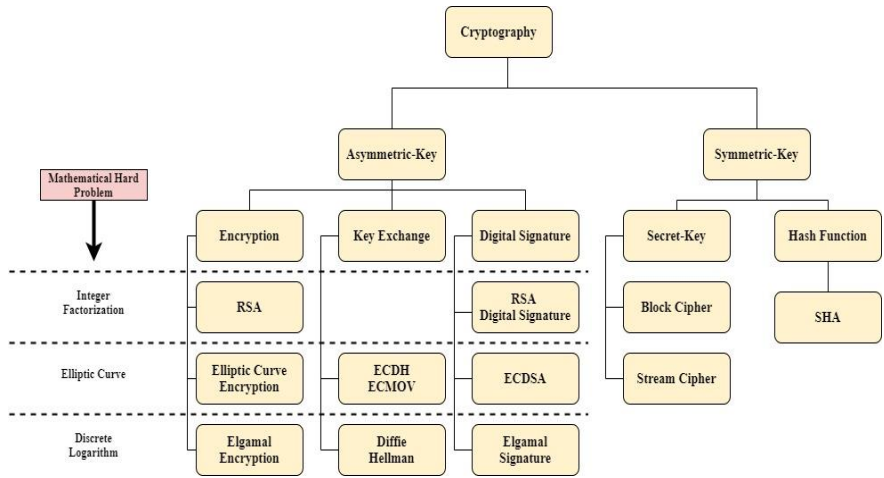

**Figure 3: -** *Cryptography branches and the associated mathematical problems*

The core strength of the RSA algorithm is based on prime factorization as a method of one-way encryption. The sender uses an encryption key which is generated using the multiplication of two randomly generated large prime numbers. The receiver can decrypt the message, only if knows the exact prime factor. The problem is easy to solve in one direction but difficult to compute in another direction. For example, multiplying the two integer numbers 8 and 18 yields 144, is a very simple calculation. However, testing all the possible combination of numbers that gives the same 144 is considered a long list (1 × 144), (2 × 72), (3 × 48), (4 × 36), (6 × 24), (8 × 18), (9 × 16) and (12 × 12). The process is slow and time-consuming for the classical computer. Most of the current cryptography algorithms rely on the assumptions that its computationally hard to break within a reasonable time.

## 4.1 The effect of an algorithm

In 1994 Peter Shor developed a polynomial-time quantum computer algorithm for integer factorization. It was a major breakthrough to prove the Quantum Computer supremacy over a classical computer [25][32]. Shor's algorithm will greatly reduce the prime factorization time. All the encryption algorithm which is based on the assumption that extracting the private key is computationally infeasible with the existing computing resource does not stand.

Successful implementation of Shor's algorithm could easily break the most widely used cryptographic algorithm such as RSA, Elliptic Curve Cryptography (ECC), and Diffie-Hellman. These algorithms play a vital role in data protection and privacy. Any possibility to break these algorithms is a serious threat and could be devastating[11][14][47]. It will be a direct threat to the information and communication security of the individuals, government and business organizations.

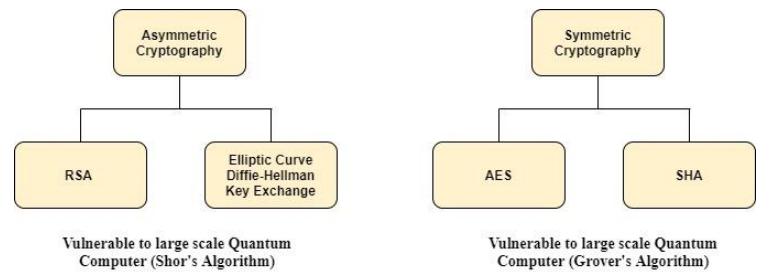

*Figure 4: -* *Effect of Shor's and Grover's algorithms on Cryptography Algorithm*

In 1996 Lov Grover developed an algorithm using quantum computers to search in the unsorted database. Grover's search algorithm can find an element in the unsorted database of N elements in $\sqrt{N}$ searches. The algorithm can search and find the key in very less iteration. For example, Grover's search algorithm needs only 185 searches to find the key for the DES algorithm which relies on a 56-bit key. Though Grover's search algorithm is not so fast as Shor's algorithm it is a potential threat for the symmetric cryptography algorithm. Collectively both Shor's algorithm and Grover's algorithm can break most of the modern cryptographic algorithms (Figure 4).

## 4.2 Sense of Urgency

As described by Dr. Michele Mosca [34], the urgency for any organization to transit to the quantum-resistant or quantum-safe cryptography depends on the following parameters:

- **Shelf-Life Time (X Years): -** Number of years you need your cryptographic key to remain secure and your data to be protected.

- **Migration Time (Y Years):** - Number of years required to develop, deploy and migrate to the Quantum-Safe solution.
- **Threat Timeline (Z Years):** - The number of years before large-scale quantum computers will be built which can break the current cryptography algorithms.

If the Threat Timeline is shorter than the sum of Shelf-Life Time and Migration Time (Figure 5) i.e. 'X+Y>Z' then it's a matter of serious concern as the organizations will not be able to protect their assets for the required years against quantum attacks[3][39][42].

Assessing the exact Threat Timeline (i.e., value for 'Z') is a challenging task as there are so many obstacles in building the quantum computer with a required number of qubits and efficiency. However, the current trend in research and development shows that the days are not far away when we will have quantum computers with the required computation power [6][13]. At some point, we can also expect Moore's Law to support the scaling up the development of quantum Computer technology as it happened with conventional computing technologies.

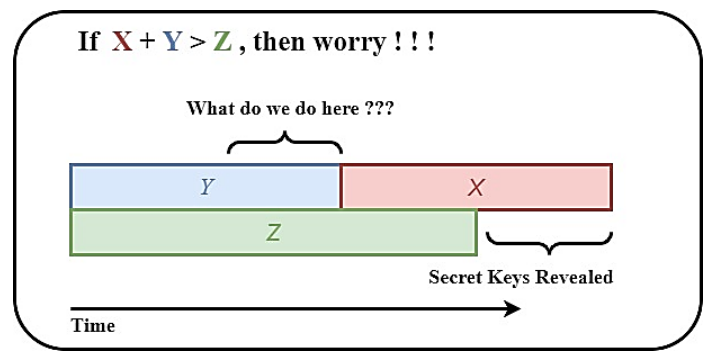

Figure 5: - Theorem (Mosca)

The threat is not only challenging for future perspective but it's significant today itself [16][57]. It's quite possible that many hackers, state-sponsored or nation-states themselves might be intercepting and harvesting the encrypted messages with envisioning that they will be able to decrypt these messages when the quantum computing resources will be available in the future. If today's private messages get disclosed even in the future, it will have a serious adverse effect on corporates, government organizations, military, and the diplomatic relations of countries. So, the threat is not in the future as shown by Mosca Theorem but has already begun from the very first day Shor's algorithm was introduced in 1994. Hence, any communication done since 1994 using an existing cryptography algorithm that is not quantum-safe is vulnerable and can be exposed in the future [4][40].

*4.3 Quantum Computing Impact on Existing Cryptography Algorithm*

A practical quantum computer with the required number of qubits (millions of qubits) will be able to break all the modern public-key cryptographic systems. National Institute of Standards and Technology (NIST) 2016 report on post-quantum cryptography shown in Table 2, highlights the impact of quantum computing on common cryptographic algorithms.

*Table 1: - Impact of quantum computing on common cryptographic algorithms*

| Cryptographic Algorithm | Type | Purpose | Impact from Large Scale Quantum Computer |
|---|---|---|---|
| AES | Symmetric Key | Encryption | Larger key sizes needed |

| SHA-2, SHA-3 | ———— | Hash Functions | Larger output needed |
|---|---|---|---|
| RSA | Public Key | Signatures, Key establishment | No longer secure |
| ECDSA, ECDH (Elliptical Curve Cryptography) | Public Key | Signatures, Key Exchange | No longer secure |
| DSA (Finite Field Cryptography) | Public Key | Signatures, Key Exchange | No longer secure |

NIST is rigorously working to analyze, test, and validate post-quantum algorithms and is expected to release a draft standard by 2023. The potential algorithms are evaluated on various factors such as robustness against cryptanalysis attack, algorithm's efficiency, interoperability, implementation feasibility, etc.

## 5. Global Initiatives and Standardization

The importance of the Quantum-Safe algorithm is well understood by government organizations and IT giants around the world. The threat is real and that day is fast approaching. Hence many countries and the standard organizations have taken an initiative for the design, development, testing, and migration strategy for quantum-safe algorithms [37]. The working group, forum, and standard organizations are publishing framework, policy documents, whitepapers, technical details, standardization documents, and cost-effective migration strategies to post-quantum cryptography. Some of the organizations leading the standardization process are: -

i) **National Institute of Standards and Technology (NIST) PQC Standardization: -** The major initiative has been taken by the National Institute of Standards and Technology (NIST) to solicit, evaluate and standardize the quantum-resistant public-key cryptographic algorithms. The NIST has begun the process in 2016 to determine the criteria and requirements for quantum-safe algorithms and announced the call for proposal [45]. The first round of submissions was announced in 2017 in which 82 submissions were received out of which 69 were announced as $1^{st}$ round candidates. Analysis for the $1^{st}$ round candidates was done in 2018 and $1^{st}$ NIST QPC standardization conference was held. The $2^{nd}$ NIST PQC standardization conference was held in the year 2019 and 26 candidates were announced for the $2^{nd}$ round. The NIST announced 7 finalists and 8 alternate candidates during the $3^{rd}$ round in the year 2020 [1].

ii) **The European Telecommunications Standards Institute (ETSI): -** The European Telecommunications Standards Institute (ETSI) has a Cyber Quantum-Safe Cryptography (QSC) group, actively working on post-quantum-safe algorithms. The QSC working groups aim to recommend a quantum-safe cryptographic algorithm and its implementation. The focus is on architectural consideration for specific applications, implementation capabilities, performance, etc. The group has published cost-effective migration strategies to post-quantum cryptography and some of the challenges and recommendations. ETSI is also organizing workshops on post-quantum cryptography since 2013.

iii) **The Internet Engineering Task Force (IETF): -** The Internet Engineering Task Force (IETF) is focused on promoting voluntary internet standards. IETF has Crypto Forum Research Group (CFRG). The CFRG forum bridge the gap between industry and research organizations, bringing new cryptographic techniques to the Internet community and promoting an understanding of the use and applicability of these mechanisms via Informational RFCs. The working group is developing Request for Comment (RFC) protocols compatible with post-quantum cryptography.

- iv) **The American National Standards Institute (ANSI): -** The Accredited Standards Committee X9F which is the subcommittee of the American National Standards Institute (ANSI) has made some recommendations on lattice-based quantum-safe algorithms as far back as 2010. The X9 committee is working actively on methods to minimize or eliminate avenues of attack by quantum computers on the financial services industry. The working group objective is to determine what preventive measures the financial service industry should be taking to protect against attacks from a large-scale quantum computer and publish periodical reports and whitepapers.
- v) **Federal Office for Information Security (BSI): -** In order to be prepared for future risk, the German Federal Office for Information Security, BSI has also released technical guidelines that recommend algorithms, key lengths and make specific recommendations of Merkle based signature schemes and XMSS. The focus is the development of quantum-safe public-key cryptography algorithms. In 2020 the BSI also developed the initial recommendations for the migration to post-quantum cryptography.

## 6. Potential Solution for Post Quantum Encryption Challenges

Scientists and research organizations are working very hard to develop and standardize quantum-safe cryptographic algorithms. However, not much information is available in the public domain except the work done by NIST. The NIST process to solicit, evaluate, and standardize the quantum-resistant cryptographic algorithms is very meticulous and the detailed information is published. Hence, decided to do the comparative analysis of 7 finalists and 8 alternate quantum-resistant algorithms announced by the NIST during the 3rd round in the year 2020. The objective of the analysis is to assess the implementation feasibility of the prospective algorithms based on their CPU cycle and memory utilization.

The performance analysis of the algorithms is done using the Open Quantum Safe (OQS) project. It's a project, developing and prototyping quantum-resistant cryptography algorithms[2]. QOS has developed "liboqs" which is an open-source library for quantum-resistant cryptographic algorithms. QOS is mainly focused on the NIST Post-Quantum Cryptography standardization for Key Encryption Mechanisms (KEM) and Signature Schemes. The OQS also provides benchmarking data for various quantum-resistant algorithms which are used in this paper for the comparative analysis of the algorithm's runtime behavior and memory consumptions. The runtime behavior and memory consumption data of the algorithms are collected based on the execution of the algorithms on Amazon Web Service (AWS) with CPU Model Intel(R) Xeon(R) Platinum 8259CL CPU @ 2.50GHz.

### *6.1 Code-Based Cryptography*

The code-based cryptography algorithm relies on the hardness of decoding in a linear error-correcting code. Robert McEliece in 1978 proposed the idea of the encryption method starts with a particular error-correcting code, a binary Goppa code, and scrambles it with an invertible linear transformation. In McEliece's idea, a public key is the product of the Goppa code and the linear transformation. To encrypt the message, the sender needs to add a specific amount of random noise [9][10]. The noise can be removed only using the Goppa code. It is a computationally challenging problem for the attacker to recover the message without knowing how to factor in the public key. There are various code-based cryptography algorithms available, among which some potential quantum-safe algorithms are: -

- i) **Classic McEliece** was introduced by Robert McEliece in 1978. It's a code-based algorithm proposed for the Key Encapsulation Mechanism (KEM). There are very few changes in the algorithm since it was introduced. The changes in the algorithm's security parameters are done to increase the computational speeds. Classic McEliece parameters can be set to match all the five NIST security levels. The computation time of the Classic McEliece algorithm is extremely fast but the algorithms require a very large public key size.

ii) **BIKE** (Bit Flipping Key Encapsulation) is a code-based key encapsulation mechanism. It is based on quasi-cyclic moderate density parity-check (QC-MDPC) codes that can be decoded using bit flipping decoding techniques. BIKE can satisfy the NIST security level 1 and 3.

iii) **HQC** (Hamming Quasi-Cyclic) is a code-based public-key algorithm. It is an IND-CPA (indistinguishability against chosen-plaintext attack) algorithm designed to provide security against classical and quantum computers. The HQC algorithm is designed to achieve NIST 1, 3, and 5 security levels.

Classic McEliece is round 3 finalists whereas BIKE, and HQC are declared as alternate candidates by NIST (Table 2). Runtime behavior and memory consumptions of Classic MeEliece, BIKE, and HQC are shown in Table 3 and Table 4.

*Table 2: - Code Based Quantum-Safe Cryptographic Algorithms*

| Sl. No. | Algorithm Name | Current Status | Open-Source C Library for Quantum-Safe Cryptographic Algorithms (liboqs) | | | | Usage |
|---|---|---|---|---|---|---|---|
| | | | NIST Level | Algorithms Name | Public Key Size (bytes) | Private Key Size (bytes) | |
| 1 | Classic McEliece | Finalists | 1 | Classic-McEliece-348864 | 261120 | 6452 | Key Encapsulation Mechanisms |
| | | | | Classic-McEliece-348864f | 261120 | 6452 | |
| | | | 3 | Classic-McEliece-460896 | 524160 | 13568 | |
| | | | | Classic-McEliece-460896f | 524160 | 13568 | |
| | | | 5 | Classic-McEliece-6688128 | 1044992 | 13892 | |
| | | | | Classic-McEliece-6688128f | 1044992 | 13892 | |
| | | | | Classic-McEliece-6960119 | 1047319 | 13908 | |
| | | | | Classic-McEliece-6960119f | 1047319 | 13908 | |
| | | | | Classic-McEliece-8192128 | 1357824 | 14080 | |
| | | | | Classic-McEliece-8192128f | 1357824 | 14080 | |
| 2 | BIKE | Alternates | 1 | BIKE-L1 | 1541 | 5223 | Key Encapsulation Mechanisms |
| | | | 3 | BIKE-L3 | 3083 | 10105 | |
| 3 | HQC | Alternates | 1 | HQC-128 | 2249 | 2289 | Key Encapsulation Mechanisms |
| | | | 3 | HQC-192 | 4522 | 4562 | |
| | | | 5 | HQC-256 | 7245 | 7285 | |

*Table 3: - Runtime Analysis of Open Quantum Safe Code Based Cryptographic Algorithms (Key Encapsulation Mechanisms)*

| Algorithm | Keygen/s | Keygen (cycles) | Encaps/s | Encaps (cycles) | Decaps/s | Decaps (cycles) |
|---|---|---|---|---|---|---|
| BIKE-L1 (x86_64) | 4256.67 | 587177 | 29602 | 84291 | 1866.67 | 1339145 |
| BIKE-L3 (x86_64) | 1345.33 | 1858211 | 14276.33 | 174889 | 558 | 4480077 |
| Classic-McEliece-348864 (x86_64) | 7.2 | 347293159 | 55104.67 | 45242 | 18448.67 | 135429 |

| Algorithm | | | | | | |
|---|---|---|---|---|---|---|
| Classic-McEliece-348864f (x86_64) | 9.24 | 270597954 | 54883 | 45426 | 17712.67 | 141044 |
| Classic-McEliece-460896 (x86_64) | 2.42 | 1032268240 | 31683.67 | 78770 | 7228.33 | 345779 |
| Classic-McEliece-460896f (x86_64) | 2.91 | 858376391 | 31720.33 | 78686 | 7220.33 | 346146 |
| Classic-McEliece-6688128 (x86_64) | 1.56 | 1604951815 | 17170 | 145425 | 6163.67 | 405515 |
| Classic-McEliece-6688128f (x86_64) | 2.2 | 1137319229 | 17034.67 | 146612 | 6112.67 | 408895 |
| Classic-McEliece-6960119 (x86_64) | 1.52 | 1647315912 | 17755.33 | 140662 | 6705.67 | 372718 |
| Classic-McEliece-6960119f (x86_64) | 2.36 | 1058000162 | 17750 | 140696 | 6655.33 | 375498 |
| Classic-McEliece-8192128 (x86_64) | 1.55 | 1617672860 | 14259.33 | 175108 | 6128 | 407881 |
| Classic-McEliece-8192128f (x86_64) | 2.13 | 1176481909 | 14550.33 | 171631 | 6146.67 | 406618 |
| HQC-128 (x86_64) | 16412.33 | 152121 | 9847.67 | 253741 | 5710.67 | 437665 |
| HQC-192 (x86_64) | 7453.33 | 335160 | 4283 | 583564 | 2422.33 | 1032052 |
| HQC-256 (x86_64) | 4232.67 | 590283 | 2430 | 1028703 | 1435.33 | 1741750 |

***Table 4:** - Memory Use Analysis (bytes per algorithm) of Open Quantum Safe Code Based Cryptographic Algorithms (Key Encapsulation Mechanisms)*

| Algorithm | Keygen (maxHeap) | Keygen (maxStack) | Encaps (maxHeap) | Encaps (maxStack) | Decaps (maxHeap) | Decaps (maxStack) |
|---|---|---|---|---|---|---|
| BIKE-L1 (x86_64) | 11420 | 90552 | 12545 | 25720 | 12577 | 73464 |
| BIKE-L3 (x86_64) | 17844 | 179448 | 20511 | 50296 | 20543 | 144952 |
| Classic-McEliece-348864 (x86_64) | 271748 | 2205032 | 276556 | 3824 | 271940 | 43232 |
| Classic-McEliece-348864f (x86_64) | 271748 | 2205032 | 276556 | 3824 | 271940 | 43232 |
| Classic-McEliece-460896 (x86_64) | 541904 | 4768360 | 546772 | 6528 | 542156 | 80256 |
| Classic-McEliece-460896f (x86_64) | 541904 | 4768360 | 546772 | 6528 | 542156 | 80256 |
| Classic-McEliece-6688128 (x86_64) | 1063060 | 4768840 | 1067980 | 6320 | 1063364 | 94144 |
| Classic-McEliece-6688128f (x86_64) | 1063060 | 4768840 | 1067980 | 6320 | 1063364 | 94144 |
| Classic-McEliece-6960119 (x86_64) | 1065403 | 4768808 | 1070309 | 6224 | 1065693 | 80896 |
| Classic-McEliece-6960119f (x86_64) | 1065403 | 4768808 | 1070309 | 6224 | 1065693 | 80896 |
| Classic-McEliece-8192128 (x86_64) | 1376080 | 4769000 | 1376080 | 384 | 1376384 | 133408 |
| Classic-McEliece-8192128f (x86_64) | 1376080 | 4769000 | 1376080 | 384 | 1376384 | 133408 |
| HQC-128 (x86_64) | 8714 | 40120 | 13259 | 53720 | 13323 | 62872 |
| HQC-192 (x86_64) | 13260 | 79640 | 22350 | 107000 | 22414 | 125304 |
| HQC-256 (x86_64) | 18706 | 182040 | 33239 | 226200 | 33303 | 255704 |

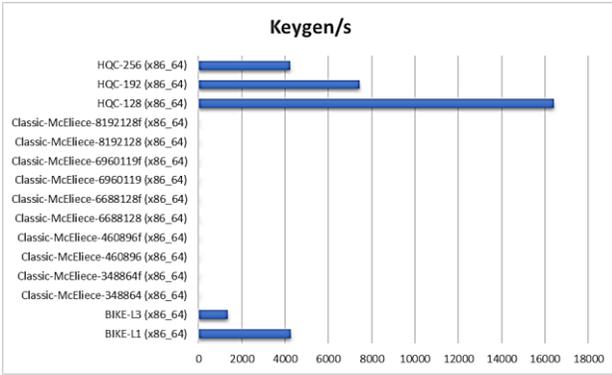

*Figure 6: - Keygen operations per second per algorithm (Key Encapsulation Mechanisms using Code Based Algorithms)*

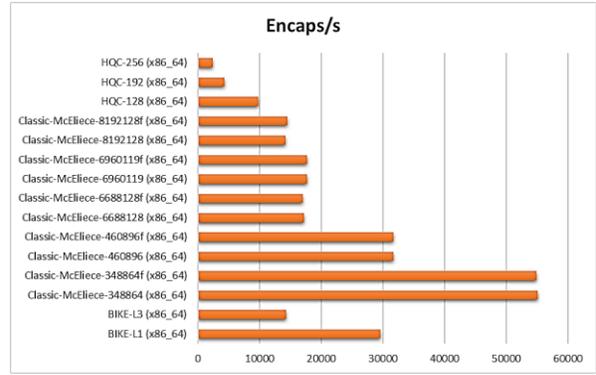

*Figure 7: - Encaps operations per second per algorithm (Key Encapsulation Mechanisms using Code Based Algorithms)*

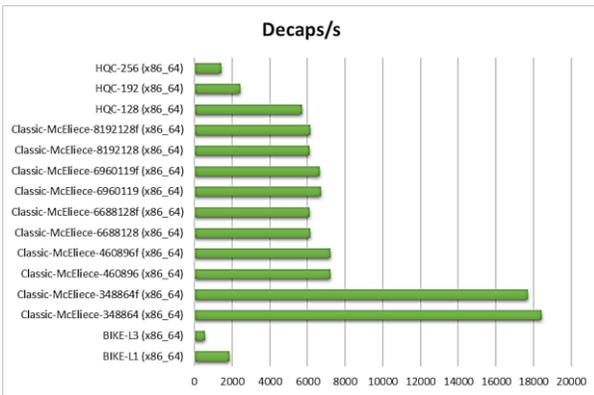

*Figure 8: - Decaps operations per second per algorithm (Key Encapsulation Mechanisms using Code Based Algorithms)*

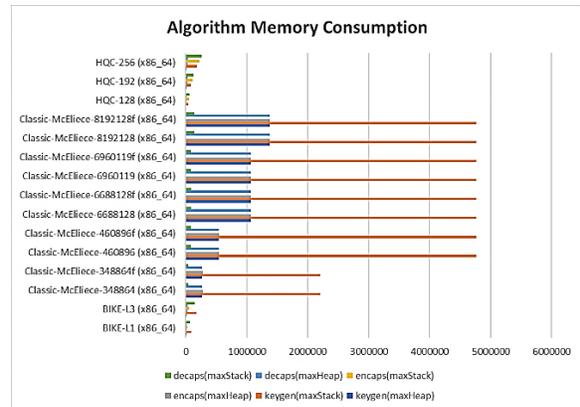

*Figure 9: - Memory consumption (bytes per algorithm) (Key Encapsulation Mechanisms using Code Based Algorithms)*

## 6.2 Multivariate Quadratic Equations Based Cryptography

A multivariate quadratic equation-based public key cryptosystem is having a set of quadratic polynomials over a finite field. In some cases, those polynomials could be defined over both grounds and as an extension field. The security of a multivariate public-key cryptosystem is based on the fact that solving the multivariate polynomial equation is proven to be NP-Hard or NP-Complete. Cryptographic algorithms based on Multivariate Quadratic Equations which are potentially quantum-safe are as follows: -

  i)  **Rainbow** algorithms belong to the family of multivariate public-key cryptosystems. It is designed by Jintai Ding and Dieter Schmidt in 2004. The algorithm is based on the Oil-Vinegar signature scheme invented by Jacques Patarin. The algorithm is based on the fact that solving a set of random multivariate quadratic systems is an NP-hard problem [46]. The algorithm can be used for digital signature and its parameters can be set to achieve NIST security levels 1, 3, and 5.

  ii) **G*e*MSS** (A Gr*eat* Multivariate Short Signature) is a multivariate-based signature scheme producing small signatures. GeMSS is based on Hidden Field Equations cryptosystem (HFE) using the minus and vinegar modifiers i.e., HFEv- [30]. The algorithm has a medium/large public key and it has a very fast verification process.

Rainbow is announced as a finalist for round 3 and GeMSS is announced as an alternate (Table 5). The runtime behavior and memory consumptions of Rainbow algorithm are shown in Table 6 and Table 7. No implementation is available for GeMSS by Open Quantum Safe Project. However, the detail performance analysis for GeMSS is available on GeMSS official website [30].

*Table 5: - Multivariate Quadratic Equations Based Quantum-Safe Cryptographic Algorithms*

| Sl. No. | Algorithm Name | Current Status | Open-Source C Library for Quantum-Safe Cryptographic Algorithms (liboqs) | | | | Usage |
|---|---|---|---|---|---|---|---|
| | | | NIST Level | Algorithms Name | Public Key Size (bytes) | Private Key Size (bytes) | |
| 1 | Rainbow | Finalists | 1 | Rainbow-I-Classic | 161600 | 103648 | Digital Signature Algorithms |
| | | | 1 | Rainbow-I-Circumzenithal | 60192 | 103648 | |
| | | | 1 | Rainbow-I-Compressed | 60192 | 64 | |
| | | | 3 | Rainbow-III-Classic | 882080 | 626048 | |
| | | | 3 | Rainbow-III-Circumzenithal | 264608 | 626048 | |
| | | | 3 | Rainbow-III-Compressed | 264608 | 64 | |
| | | | 5 | Rainbow-V-Classic | 1930600 | 1408736 | |
| | | | 5 | Rainbow-V-Circumzenithal | 536136 | 1408736 | |
| | | | 5 | Rainbow-V-Compressed | 536136 | 64 | |
| 2 | GeMSS | Alternates | 1 | N/A (No implementation by Open Quantum Safe Project) | 352,180 | 32 | Digital Signature Algorithms |
| | | | 3 | | | | |
| | | | 5 | | | | |

*Table 6: - Runtime Analysis of Open Quantum Safe Multivariate Quadratic Equations Based Cryptographic Algorithms (Digital Signature Algorithms)*

| Algorithm | Keypair/s | Keypair (cycles) | Sign/s | Sign (cycles) | Verify/s | Verify (cycles) |
|---|---|---|---|---|---|---|
| Rainbow-I-Circumzenithal (x86_64) | 6.06 | 412509871 | 530.3 | 4714048 | 445.7 | 5609619 |
| Rainbow-I-Classic (x86_64) | 6.77 | 369413745 | 506 | 4940751 | 500.7 | 4993923 |
| Rainbow-I-Compressed (x86_64) | 6.02 | 415229320 | 13.77 | 181591225 | 445.9 | 5607001 |
| Rainbow-III-Circumzenithal (x86_64) | 0.43 | 1535117154 | 54.95 | 45496023 | 49.06 | 50968436 |
| Rainbow-III-Classic (x86_64) | 0.49 | 815304452 | 54.82 | 45599192 | 50.58 | 49426107 |
| Rainbow-III-Compressed (x86_64) | 0.43 | 1526552696 | 0.92 | 2709509453 | 49.2 | 50814366 |
| Rainbow-V-Circumzenithal (x86_64) | 0.15 | 3612623138 | 24.87 | 100519687 | 22.36 | 111796728 |
| Rainbow-V-Classic (x86_64) | 0.17 | 1503271522 | 23.78 | 105150930 | 24.24 | 103104686 |
| Rainbow-V-Compressed (x86_64) | 0.15 | 3653272908 | 0.32 | 3503174534 | 21.43 | 116636792 |

*Table 7: - Memory Use Analysis (bytes per algorithm) of Open Quantum Safe Multivariate Quadratic Equations Based Cryptographic Algorithms (Digital Signature Algorithms)*

| Algorithm | Keygen (maxHeap) | Keygen (maxStack) | Sign (maxHeap) | Sign (maxStack) | Verify (maxHeap) | Verify (maxStack) |
|---|---|---|---|---|---|---|
| Rainbow-I-Circumzenithal (x86_64) | 168008 | 142440 | 172822 | 912 | 168174 | 324056 |
| Rainbow-I-Classic (x86_64) | 269416 | 178680 | 274230 | 912 | 274230 | 912 |
| Rainbow-I-Compressed (x86_64) | 64424 | 246104 | 64590 | 224504 | 64590 | 324056 |
| Rainbow-III-Circumzenithal (x86_64) | 894824 | 946696 | 895088 | 13464 | 895088 | 1765144 |
| Rainbow-III-Classic (x86_64) | 1512296 | 993416 | 1512832 | 15008 | 1512296 | 384 |
| Rainbow-III-Compressed (x86_64) | 268840 | 1572760 | 269104 | 1302104 | 269104 | 1765144 |
| Rainbow-V-Circumzenithal (x86_64) | 1949040 | 2208248 | 1949352 | 22024 | 1949352 | 3862296 |
| Rainbow-V-Classic (x86_64) | 3343504 | 2193512 | 3343504 | 384 | 3343504 | 384 |
| Rainbow-V-Compressed (x86_64) | 540368 | 3617000 | 540680 | 2901304 | 540680 | 3862296 |

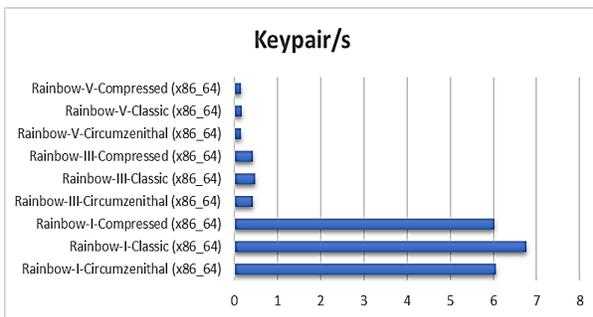

*Figure 10: - Keypair operations per second per algorithm (Digital Signature using Multivariate Quadratic Equations Based Algorithms)*

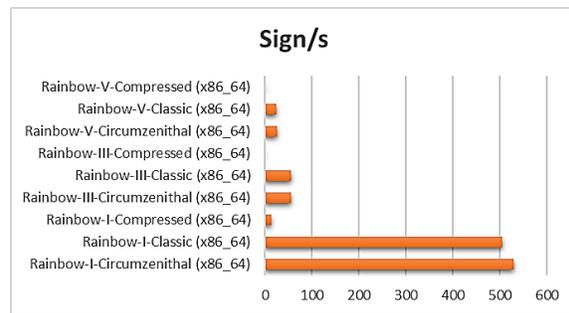

*Figure 11: - Sign operations per second per algorithm (Digital Signature using Multivariate Quadratic Equations Based Algorithms)*

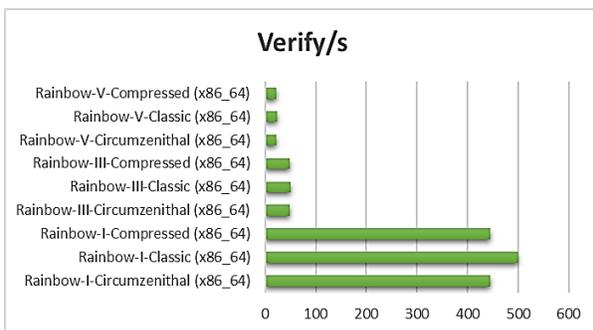

*Figure 12: - Verify operations per second per algorithm (Digital Signature using Multivariate Quadratic Equations Based Algorithms)*

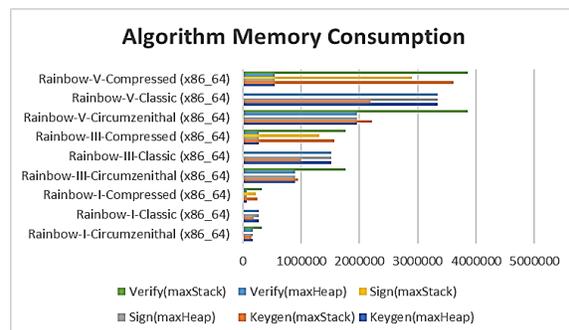

*Figure 13: - Memory consumption (bytes per algorithm) (Digital Signature using Multivariate Quadratic Equations Based Algorithms)*

## 6.3 Hash-Based Cryptography

Cryptographic hash functions are non-reversible function that takes a string of any length as input and produces a fixed-length output. Hash-based cryptography is popularly used for digital signatures such as Merkle signature. It combines a one-time signature with the Merkle tree. The hash-based signature (HBS) is a collection of many one-time signature (OTS) schemes. HBS uses a tree data structure to combine many OTS in an efficient manner. To sign a message, an HBS chooses a single OTS from its collection and uses it to sign the message [20]. The critical issue is that the HBS must never choose the same OTS twice, or else security is compromised. Hash-based quantum-safe cryptographic algorithms are: -

i) **SPHINCS+** is a stateless hash-based signature scheme. It is an improvised version of SPHINCS, especially aimed to reduce the signature size [51]. The NIST submission of SPHINCS+ proposes three different signature schemes:

   o SPHINCS+-SHAKE256
   o SPHINCS+-SHA-256
   o SPHINCS+-Haraka

SPHINCS+ is announced as an alternate by NIST in the 3rd round of algorithm selection (Table 8). The algorithm can be used for digital signature and can satisfy the NIST 1, 3, and 5 security levels. Runtime behavior and memory consumptions of SPHINCA+ are shown in Table 9 and Table 10.

*Table 8: - Hash-Based Quantum-Safe Cryptographic Algorithms*

| Sl. No. | Algorithm Name | Current Status | Open-Source C Library for Quantum-Safe Cryptographic Algorithms (liboqs) | | | | Usage |
|---|---|---|---|---|---|---|---|
| | | | NIST Level | Algorithms Name | Public Key Size (bytes) | Private Key Size (bytes) | |
| 1 | SPHINCS+ | Alternates | 1 | SPHINCS+-Haraka-128f-robust | 32 | 64 | Digital Signature Algorithms |
| | | | | SPHINCS+-Haraka-128f-simple | 32 | 64 | |
| | | | | SPHINCS+-Haraka-128s-robust | 32 | 64 | |
| | | | | SPHINCS+-Haraka-128s-simple | 32 | 64 | |
| | | | | SPHINCS+-SHA256-128f-robust | 32 | 64 | |
| | | | | SPHINCS+-SHA256-128f-simple | 32 | 64 | |
| | | | | SPHINCS+-SHA256-128s-robust | 32 | 64 | |
| | | | | SPHINCS+-SHA256-128s-simple | 32 | 64 | |
| | | | | SPHINCS+-SHAKE256-128f-robust | 32 | 64 | |

| | | | | SPHINCS+-SHAKE256-128f-simple | 32 | 64 | |
|---|---|---|---|---|---|---|---|
| | | | | SPHINCS+-SHAKE256-128s-robust | 32 | 64 | |
| | | | | SPHINCS+-SHAKE256-128s-simple | 32 | 64 | |
| | | | 3 | SPHINCS+-Haraka-192f-robust | 48 | 96 | |
| | | | | SPHINCS+-Haraka-192f-simple | 48 | 96 | |
| | | | | SPHINCS+-Haraka-192s-robust | 48 | 96 | |
| | | | | SPHINCS+-Haraka-192s-simple | 48 | 96 | |
| | | | | SPHINCS+-SHA256-192f-robust | 48 | 96 | |
| | | | | SPHINCS+-SHA256-192f-simple | 48 | 96 | |
| | | | | SPHINCS+-SHA256-192s-robust | 48 | 96 | |
| | | | | SPHINCS+-SHA256-192s-simple | 48 | 96 | |
| | | | | SPHINCS+-SHAKE256-192f-robust | 48 | 96 | |
| | | | | SPHINCS+-SHAKE256-192f-simple | 48 | 96 | |
| | | | | SPHINCS+-SHAKE256-192s-robust | 48 | 96 | |
| | | | | SPHINCS+-SHAKE256-192s-simple | 48 | 96 | |
| | | | 5 | SPHINCS+-Haraka-256f-robust | 64 | 128 | |
| | | | | SPHINCS+-Haraka-256f-simple | 64 | 128 | |
| | | | | SPHINCS+-Haraka-256s-robust | 64 | 128 | |
| | | | | SPHINCS+-Haraka-256s-simple | 64 | 128 | |
| | | | | SPHINCS+-SHA256-256f-robust | 64 | 128 | |
| | | | | SPHINCS+-SHA256-256f-simple | 64 | 128 | |

| | | | SPHINCS+-SHA256-256s-robust | 64 | 128 | |
| | | | SPHINCS+-SHA256-256s-simple | 64 | 128 | |
| | | | SPHINCS+-SHAKE256-256f-robust | 64 | 128 | |
| | | | SPHINCS+-SHAKE256-256f-simple | 64 | 128 | |
| | | | SPHINCS+-SHAKE256-256s-robust | 64 | 128 | |
| | | | SPHINCS+-SHAKE256-256s-simple | 64 | 128 | |

*Table 9: - Runtime Analysis of Open Quantum Safe Hash Based Cryptographic Algorithms (Digital Signature Algorithms)*

| Algorithm | Keypair/s | Keypair (cycles) | Sign/s | Sign (cycles) | Verify/s | Verify (cycles) |
|---|---|---|---|---|---|---|
| SPHINCS+-Haraka-128f-robust (x86_64) | 3305.33 | 756217 | 130.25 | 19193087 | 1841.3 | 1357668 |
| SPHINCS+-Haraka-128f-simple (x86_64) | 3021.67 | 827241 | 126.75 | 19720897 | 2389.7 | 1046180 |
| SPHINCS+-Haraka-128s-robust (x86_64) | 49.8 | 50199666 | 6.26 | 399193532 | 5109 | 489226 |
| SPHINCS+-Haraka-128s-simple (x86_64) | 53.95 | 46336950 | 6.88 | 363156129 | 7042 | 354926 |
| SPHINCS+-Haraka-192f-robust (x86_64) | 2491 | 1003522 | 78.92 | 31672906 | 1248 | 2003318 |
| SPHINCS+-Haraka-192f-simple (x86_64) | 2850 | 877072 | 99.34 | 25163368 | 1754.7 | 1424860 |
| SPHINCS+-Haraka-192s-robust (x86_64) | 26.88 | 92991222 | 2.63 | 950449292 | 2971.7 | 841268 |
| SPHINCS+-Haraka-192s-simple (x86_64) | 30.68 | 81486323 | 3.15 | 794443486 | 3977.3 | 628454 |
| SPHINCS+-Haraka-256f-robust (x86_64) | 802.67 | 3114999 | 36.18 | 69108207 | 1203.6 | 2076672 |
| SPHINCS+-Haraka-256f-simple (x86_64) | 823.73 | 3035014 | 38.42 | 65055419 | 1592.1 | 1569822 |
| SPHINCS+-Haraka-256s-robust (x86_64) | 47.11 | 53060953 | 3.34 | 748586944 | 2257.3 | 1107434 |
| SPHINCS+-Haraka-256s-simple (x86_64) | 51.25 | 48787641 | 3.93 | 636871172 | 3061.7 | 816421 |
| SPHINCS+-SHA256-128f-robust (x86_64) | 962.35 | 2597396 | 34.83 | 71789115 | 216.64 | 11537909 |
| SPHINCS+-SHA256-128f-simple (x86_64) | 1833.33 | 1363722 | 65.69 | 38055893 | 410.73 | 6087008 |
| SPHINCS+-SHA256-128s-robust (x86_64) | 15.42 | 162114391 | 1.94 | 1289794883 | 574.14 | 4354738 |
| SPHINCS+-SHA256-128s-simple (x86_64) | 27.71 | 90208011 | 3.64 | 687556446 | 1228.3 | 2035202 |
| SPHINCS+-SHA256-192f-robust (x86_64) | 676.11 | 3697467 | 22.13 | 112975498 | 144.19 | 17338755 |
| SPHINCS+-SHA256-192f-simple (x86_64) | 1200.33 | 2082895 | 38.49 | 64952624 | 250.42 | 9983430 |
| SPHINCS+-SHA256-192s-robust (x86_64) | 10.41 | 240149708 | 1.08 | 2317742108 | 399.4 | 6258715 |
| SPHINCS+-SHA256-192s-simple (x86_64) | 18.74 | 133365900 | 1.88 | 1331847740 | 730.33 | 3423401 |

| Algorithm | | | | | | |
|---|---|---|---|---|---|---|
| SPHINCS+-SHA256-256f-robust (x86_64) | 140.24 | 17824497 | 6.67 | 374615844 | 108.67 | 23007612 |
| SPHINCS+-SHA256-256f-simple (x86_64) | 450.85 | 5544896 | 20.81 | 120145814 | 272.91 | 9159588 |
| SPHINCS+-SHA256-256s-robust (x86_64) | 8.73 | 286251953 | 0.75 | 3329041447 | 207.26 | 12062042 |
| SPHINCS+-SHA256-256s-simple (x86_64) | 30.34 | 82391795 | 2.4 | 1043537732 | 530.33 | 4714332 |
| SPHINCS+-SHAKE256-128f-robust (x86_64) | 462.36 | 5406859 | 18.56 | 134699425 | 154.69 | 16161667 |
| SPHINCS+-SHAKE256-128f-simple (x86_64) | 844 | 2962241 | 32.16 | 77738565 | 301.8 | 8283439 |
| SPHINCS+-SHAKE256-128s-robust (x86_64) | 7.81 | 320032253 | 1.02 | 2460642112 | 543.49 | 4599201 |
| SPHINCS+-SHAKE256-128s-simple (x86_64) | 13.66 | 183072418 | 1.71 | 1463603275 | 915.67 | 2730203 |
| SPHINCS+-SHAKE256-192f-robust (x86_64) | 331.89 | 7532419 | 11.55 | 216435158 | 119.17 | 20977226 |
| SPHINCS+-SHAKE256-192f-simple (x86_64) | 580.14 | 4309068 | 20.11 | 124327976 | 229.26 | 10905347 |
| SPHINCS+-SHAKE256-192s-robust (x86_64) | 5.2 | 480506759 | 0.58 | 46486796 | 319.23 | 7831342 |
| SPHINCS+-SHAKE256-192s-simple (x86_64) | 9.35 | 267514170 | 1 | 2498600938 | 655.33 | 3814860 |
| SPHINCS+-SHAKE256-256f-robust (x86_64) | 123.63 | 20220286 | 5.92 | 422251215 | 108.78 | 22981847 |
| SPHINCS+-SHAKE256-256f-simple (x86_64) | 220.11 | 11357895 | 10.37 | 241136741 | 209.39 | 11937656 |
| SPHINCS+-SHAKE256-256s-robust (x86_64) | 7.5 | 333331669 | 0.66 | 3783521422 | 227.85 | 10973738 |
| SPHINCS+-SHAKE256-256s-simple (x86_64) | 13.28 | 188315722 | 1.08 | 2311104060 | 427.19 | 5852661 |

*Table 10: -Memory Use Analysis (bytes per algorithm) of Open Quantum Safe Hash-Based Cryptographic Algorithms (Digital Signature Algorithms)*

| Algorithm | Keygen (maxHeap) | Keygen (maxStack) | Sign (maxHeap) | Sign (maxStack) | Verify (maxHeap) | Verify (maxStack) |
|---|---|---|---|---|---|---|
| SPHINCS+-Haraka-128f-robust (x86_64) | 8912 | 3712 | 26100 | 19472 | 26100 | 2304 |
| SPHINCS+-Haraka-128f-simple (x86_64) | 8912 | 3712 | 26100 | 19472 | 26100 | 2304 |
| SPHINCS+-Haraka-128s-robust (x86_64) | 8912 | 2368 | 16868 | 10240 | 16868 | 2304 |
| SPHINCS+-Haraka-128s-simple (x86_64) | 8912 | 2368 | 16868 | 10240 | 16868 | 2304 |
| SPHINCS+-Haraka-192f-robust (x86_64) | 8960 | 2432 | 44724 | 38128 | 44724 | 1960 |
| SPHINCS+-Haraka-192f-simple (x86_64) | 8960 | 2432 | 44724 | 38128 | 44724 | 1960 |
| SPHINCS+-Haraka-192s-robust (x86_64) | 8960 | 2432 | 25284 | 18672 | 25284 | 2304 |
| SPHINCS+-Haraka-192s-simple (x86_64) | 8960 | 2432 | 25284 | 18672 | 25284 | 2304 |
| SPHINCS+-Haraka-256f-robust (x86_64) | 9008 | 2496 | 58964 | 52352 | 58964 | 1960 |
| SPHINCS+-Haraka-256f-simple (x86_64) | 9008 | 2496 | 58964 | 52352 | 58964 | 1960 |
| SPHINCS+-Haraka-256s-robust (x86_64) | 4360 | 7400 | 34252 | 38552 | 38900 | 1960 |
| SPHINCS+-Haraka-256s-simple (x86_64) | 4360 | 7400 | 34252 | 37800 | 38900 | 1960 |
| SPHINCS+-SHA256-128f-robust (x86_64) | 4432 | 9416 | 21620 | 11752 | 26100 | 2304 |

| Algorithm | | | | | | |
|---|---|---|---|---|---|---|
| SPHINCS+-SHA256-128f-simple (x86_64) | 4432 | 7560 | 21620 | 9928 | 26100 | 2304 |
| SPHINCS+-SHA256-128s-robust (x86_64) | 4432 | 9640 | 12388 | 12840 | 16868 | 2304 |
| SPHINCS+-SHA256-128s-simple (x86_64) | 4432 | 7784 | 12388 | 11016 | 16868 | 2304 |
| SPHINCS+-SHA256-192f-robust (x86_64) | 4480 | 10536 | 40244 | 14120 | 40412 | 6816 |
| SPHINCS+-SHA256-192f-simple (x86_64) | 4480 | 8488 | 40244 | 12072 | 44724 | 1960 |
| SPHINCS+-SHA256-192s-robust (x86_64) | 4480 | 10856 | 20804 | 15944 | 25284 | 2304 |
| SPHINCS+-SHA256-192s-simple (x86_64) | 4480 | 8808 | 20804 | 13896 | 25284 | 2304 |
| SPHINCS+-SHA256-256f-robust (x86_64) | 4528 | 11944 | 54484 | 16360 | 54652 | 7840 |
| SPHINCS+-SHA256-256f-simple (x86_64) | 4528 | 9704 | 54484 | 14152 | 54652 | 7840 |
| SPHINCS+-SHA256-256s-robust (x86_64) | 4528 | 12200 | 34420 | 18376 | 34588 | 7840 |
| SPHINCS+-SHA256-256s-simple (x86_64) | 4528 | 9992 | 34420 | 16168 | 34588 | 6720 |
| SPHINCS+-SHAKE256-128f-robust (x86_64) | 8912 | 2432 | 26100 | 18192 | 26100 | 2304 |
| SPHINCS+-SHAKE256-128f-simple (x86_64) | 8912 | 2432 | 26100 | 18192 | 26100 | 2304 |
| SPHINCS+-SHAKE256-128s-robust (x86_64) | 8912 | 2560 | 13052 | 13128 | 16868 | 2304 |
| SPHINCS+-SHAKE256-128s-simple (x86_64) | 8912 | 2560 | 13052 | 13064 | 16868 | 2304 |
| SPHINCS+-SHAKE256-192f-robust (x86_64) | 8960 | 2432 | 40908 | 41608 | 44724 | 1960 |
| SPHINCS+-SHAKE256-192f-simple (x86_64) | 8960 | 2432 | 40908 | 41608 | 44724 | 1960 |
| SPHINCS+-SHAKE256-192s-robust (x86_64) | 8960 | 2432 | 21468 | 22888 | 25284 | 2304 |
| SPHINCS+-SHAKE256-192s-simple (x86_64) | 8960 | 2432 | 21468 | 22888 | 25284 | 2304 |
| SPHINCS+-SHAKE256-256f-robust (x86_64) | 9008 | 2496 | 55148 | 57000 | 58964 | 1960 |
| SPHINCS+-SHAKE256-256f-simple (x86_64) | 9008 | 2496 | 55148 | 57000 | 58964 | 1960 |
| SPHINCS+-SHAKE256-256s-robust (x86_64) | 5192 | 6464 | 35084 | 37800 | 38900 | 1960 |
| SPHINCS+-SHAKE256-256s-simple (x86_64) | 5192 | 6464 | 35084 | 37800 | 38900 | 1960 |

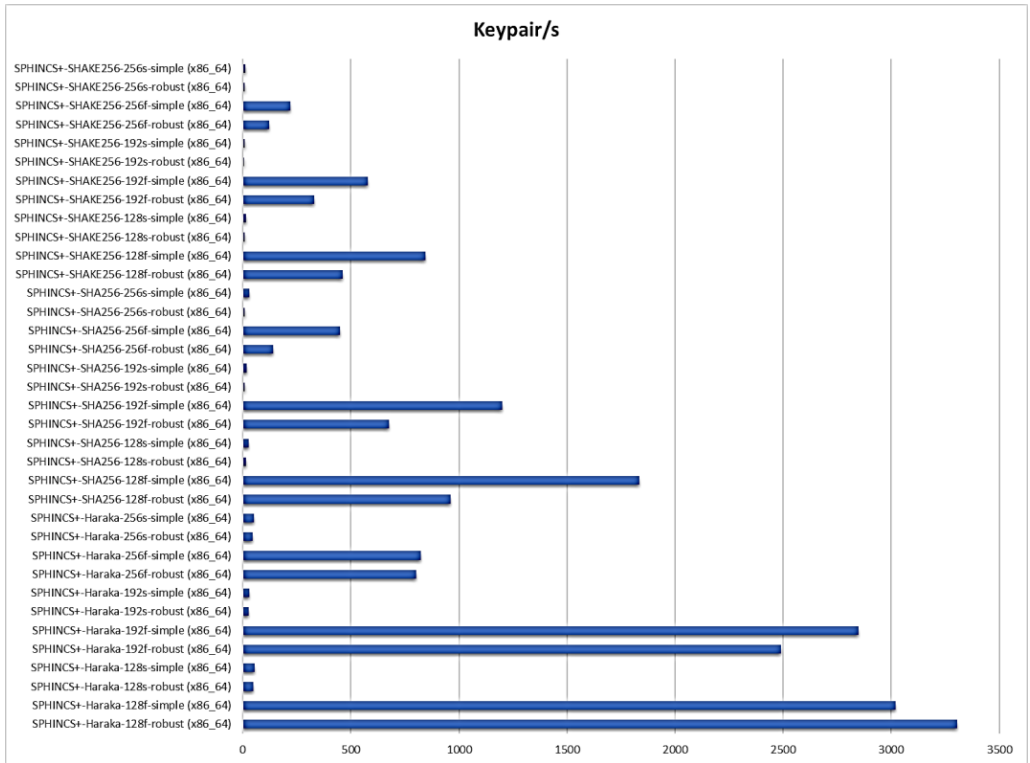

*Figure 14: -* *Keypair operations per second per algorithm (Digital Signature using Hash-Based Algorithms)*

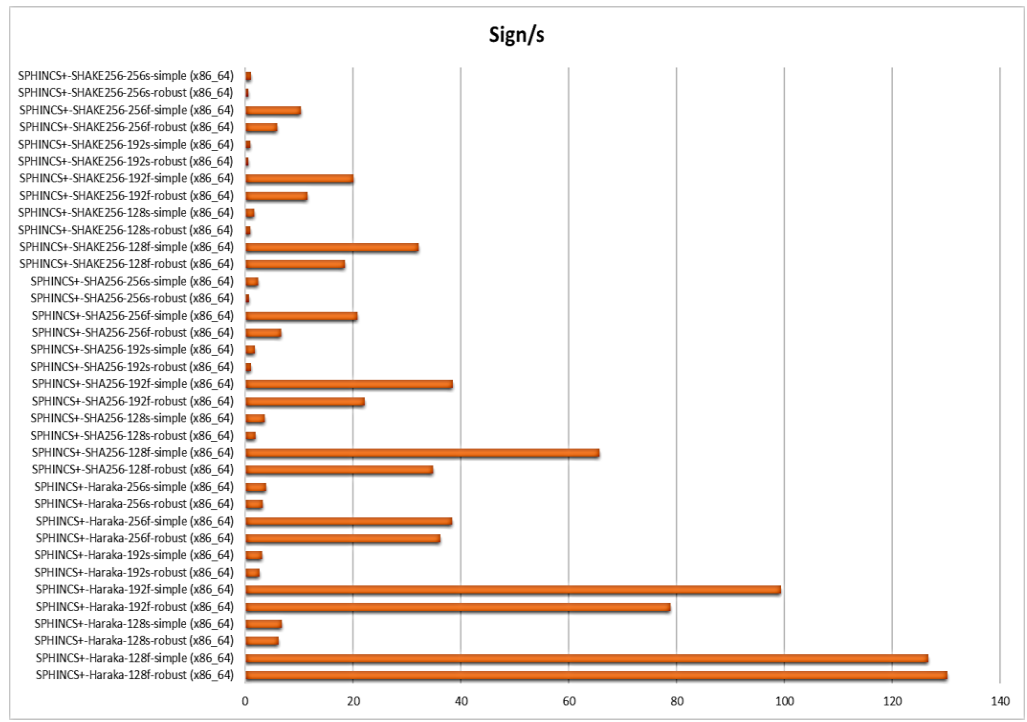

*Figure 15: -* *Sign operations per second per algorithm (Digital Signature using Hash-Based Algorithms)*

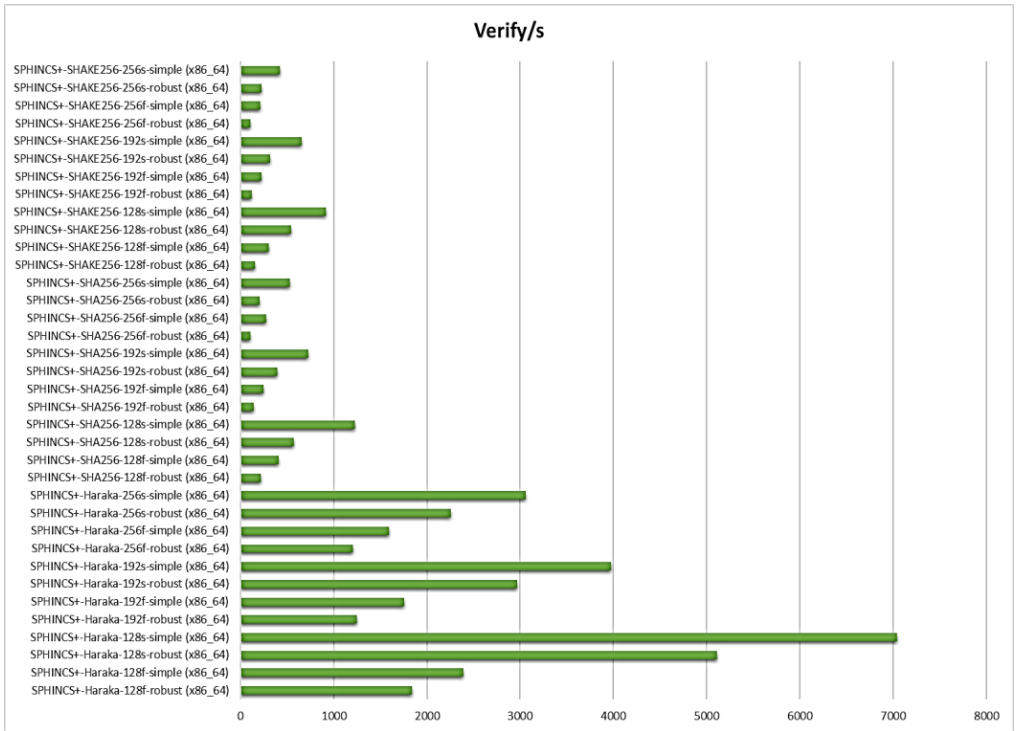

*Figure 16: - Verify operations per second per algorithm (Digital Signature using Hash-Based Algorithms)*

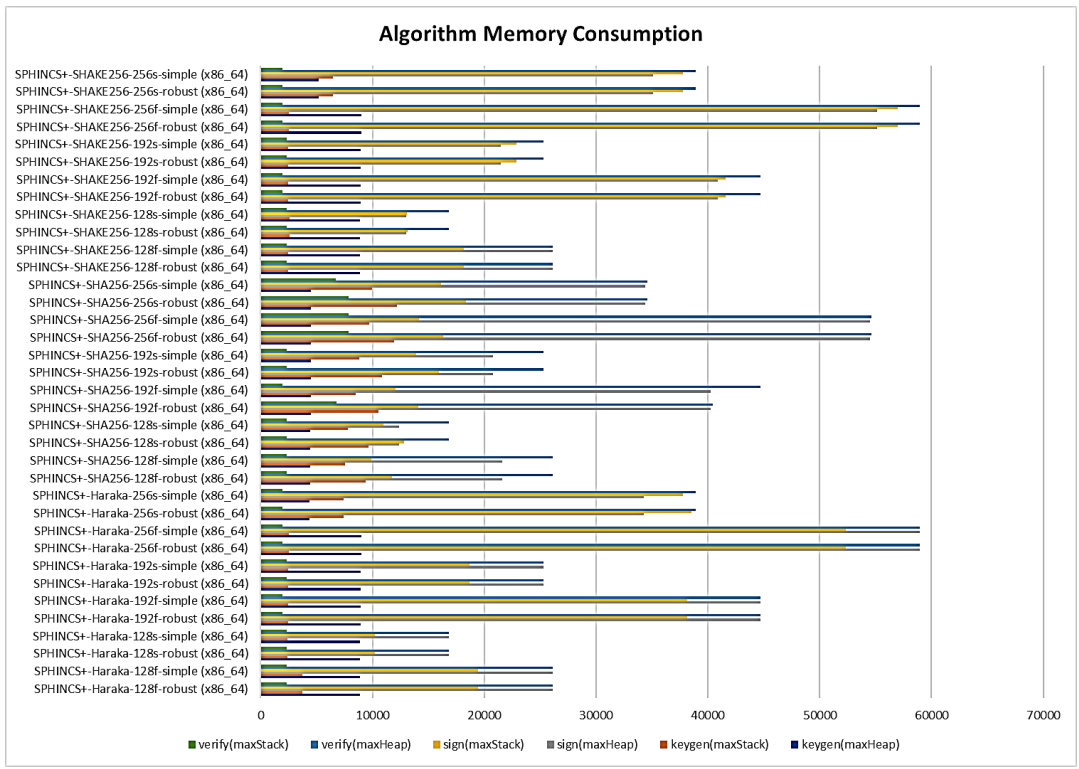

*Figure 17: - Memory consumption (bytes per algorithm) (Digital Signature using Hash-Based Algorithms)*

## 6.4 Isogeny Based Cryptography

It is a relatively new technique that has begun in the year 2000. Isogeny-based cryptography uses maps between elliptic curves to build public-key cryptosystems. The security of Isogeny cryptography is based on so-called supersingular isogeny problems or in other words finding the isogeny mapping between two supersingular elliptic curves with the same number of points. The Isogeny-based protocols require a very small key compared to any other post-quantum cryptography candidates. SIKE is the one potential quantum-safe algorithm belonging to Isogeny cryptography family [21][27][52].

i) **SIKE** (Supersingular Isogeny Key Encapsulation) is an isogeny-based key encapsulation algorithm. The algorithm implementation is based on pseudo-random walks in supersingular isogeny graphs. SIKE has the smallest public key size but it's also one of the slowest algorithms [7][50].

SIKE is the only one algorithm based on Isogney, which could make the place in the NIST 3rd round (Table 11). However, it is declared as an alternate candidate, not the finalist. The runtime behavior and memory consumptions of the algorithm are shown in Table 12 and Table 13.

*Table 11: - Isogeny Based Quantum-Safe Cryptographic Algorithms*

| Sl. No. | Algorithm Name | Current Status | Open-Source C Library for Quantum-Safe Cryptographic Algorithms (liboqs) | | | | Usage |
|---|---|---|---|---|---|---|---|
| | | | NIST Level | Algorithms Name | Public Key Size (bytes) | Private Key Size (bytes) | |
| 1 | SIKE | Alternates | 1 | SIDH-p434 | 330 | 28 | Key Encapsulation Mechanisms |
| | | | | SIDH-p434-compressed | 197 | 28 | |
| | | | | SIKE-p434 | 330 | 374 | |
| | | | | SIKE-p434-compressed | 197 | 350 | |
| | | | 2 | SIDH-p503 | 378 | 32 | |
| | | | | SIDH-p503-compressed | 225 | 32 | |
| | | | | SIKE-p503 | 378 | 434 | |
| | | | | SIKE-p503-compressed | 225 | 407 | |
| | | | 3 | SIDH-p610 | 462 | 39 | |
| | | | | SIDH-p610-compressed | 274 | 39 | |
| | | | | SIKE-p610 | 462 | 524 | |
| | | | | SIKE-p610-compressed | 274 | 491 | |
| | | | 5 | SIDH-p75 | 564 | 48 | |
| | | | | SIDH-p751-compressed | 335 | 48 | |
| | | | | SIKE-p751 | 564 | 644 | |
| | | | | SIKE-p751-compressed | 335 | 602 | |

*Table 12: -Runtime Analysis of Open Quantum Safe Isogeny Based Cryptographic Algorithms (Key Encapsulation Mechanisms)*

| Algorithm | Keygen/s | Keygen (cycles) | Encaps/s | Encaps (cycles) | Decaps/s | Decaps (cycles) |
|---|---|---|---|---|---|---|
| SIDH-p434 (x86_64) | 109.63 | 22805106 | 53.61 | 46626958 | 135.2 | 18488925 |
| SIDH-p434-compressed (x86_64) | 57.09 | 43797293 | 38.69 | 64616727 | 119.7 | 20885902 |
| SIDH-p503 (x86_64) | 318.89 | 7840371 | 155.4 | 16085706 | 391.2 | 6389639 |
| SIDH-p503-compressed (x86_64) | 166.44 | 15020665 | 111.7 | 22385569 | 330.9 | 7555453 |
| SIDH-p610 (x86_64) | 30.91 | 80887652 | 15.93 | 156909224 | 37.1 | 67391175 |
| SIDH-p610-compressed (x86_64) | 18.88 | 132403975 | 13.28 | 188156935 | 36.07 | 69302679 |
| SIDH-p751 (x86_64) | 104.79 | 23858838 | 50.17 | 49829353 | 127.4 | 19629441 |
| SIDH-p751-compressed (x86_64) | 55.93 | 44696733 | 36.57 | 68372097 | 114.3 | 21873987 |
| SIKE-p434 (x86_64) | 98.8 | 25300016 | 60.57 | 41275416 | 56.52 | 44237605 |
| SIKE-p434-compressed (x86_64) | 57.71 | 43322569 | 36.58 | 68330870 | 52.79 | 47358432 |
| SIKE-p503 (x86_64) | 271.82 | 9197872 | 175 | 14283364 | 163 | 15339004 |
| SIKE-p503-compressed (x86_64) | 165.89 | 15070206 | 112.1 | 22309024 | 154.3 | 16200750 |
| SIKE-p610 (x86_64) | 30.84 | 81071205 | 16.84 | 148430862 | 16.75 | 149271935 |
| SIKE-p610-compressed (x86_64) | 18.99 | 131620069 | 13.28 | 188252832 | 16.99 | 147137618 |
| SIKE-p751 (x86_64) | 92.91 | 26906560 | 57.49 | 43480834 | 53.45 | 46777582 |
| SIKE-p751-compressed (x86_64) | 55.8 | 44799410 | 36.53 | 68434047 | 47.73 | 52377374 |

*Table 13: -Memory Use Analysis (bytes per algorithm) of Open Quantum Safe Isogeny Based Cryptographic Algorithms (Key Encapsulation Mechanisms)*

| Algorithm | Keygen (maxHeap) | Keygen (maxStack) | Encaps (maxHeap) | Encaps (maxStack) | Decaps (maxHeap) | Decaps (maxStack) |
|---|---|---|---|---|---|---|
| SIDH-p434 (x86_64) | 4534 | 9928 | 5002 | 10376 | 5084 | 9576 |
| SIDH-p434-compressed (x86_64) | 4401 | 78248 | 4736 | 57624 | 4818 | 9576 |
| SIDH-p503 (x86_64) | 9234 | 2192 | 9738 | 2320 | 9864 | 2320 |
| SIDH-p503-compressed (x86_64) | 4433 | 92456 | 4816 | 63840 | 9558 | 2304 |
| SIDH-p610 (x86_64) | 4677 | 14520 | 5332 | 15112 | 5447 | 14008 |
| SIDH-p610-compressed (x86_64) | 4489 | 146840 | 4956 | 104072 | 5071 | 15352 |
| SIDH-p751 (x86_64) | 4788 | 11416 | 5588 | 12216 | 5728 | 10440 |
| SIDH-p751-compressed (x86_64) | 4559 | 196440 | 5130 | 123944 | 5270 | 10760 |
| SIKE-p434 (x86_64) | 4880 | 10152 | 5242 | 10504 | 5258 | 11192 |
| SIKE-p434-compressed (x86_64) | 4723 | 78248 | 4975 | 58072 | 4991 | 15912 |
| SIKE-p503 (x86_64) | 9636 | 2192 | 10062 | 2832 | 5438 | 7768 |
| SIKE-p503-compressed (x86_64) | 4808 | 92456 | 5112 | 64368 | 5136 | 13784 |
| SIKE-p610 (x86_64) | 5162 | 14856 | 5672 | 15320 | 5696 | 16248 |
| SIKE-p610-compressed (x86_64) | 4941 | 146840 | 5301 | 104712 | 5325 | 25144 |
| SIKE-p751 (x86_64) | 5384 | 12200 | 6012 | 12376 | 6044 | 12792 |
| SIKE-p751-compressed (x86_64) | 5113 | 196456 | 5555 | 124696 | 5587 | 22440 |

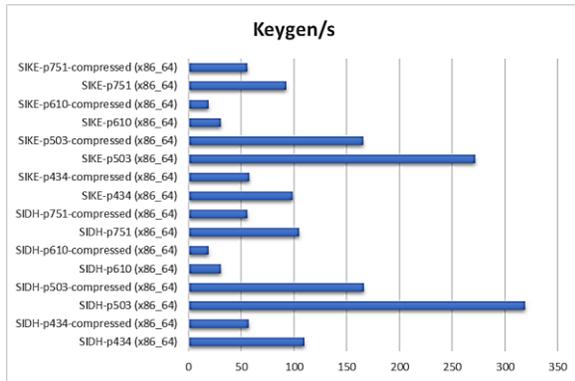

*Figure 18: -* *Keygen operations per second per algorithm (Key Encapsulation Mechanisms using Isogeny Based Algorithms)*

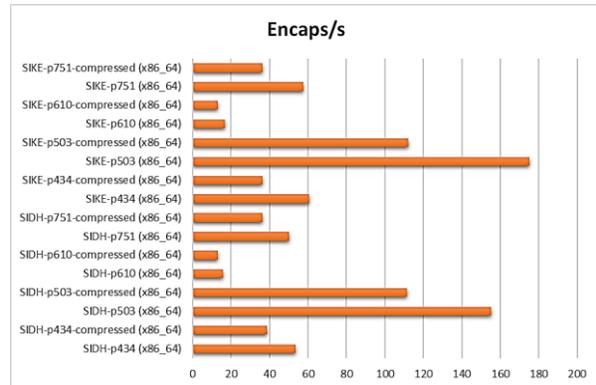

*Figure 19: -* *Encaps operations per second per algorithm (Key Encapsulation Mechanisms using Isogeny Based Algorithms)*

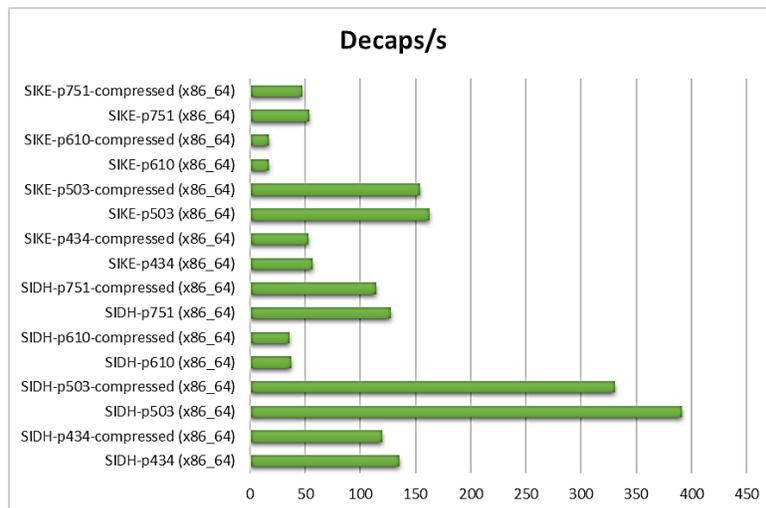

*Figure 20: -* *Decaps operations per second per algorithm (Key Encapsulation Mechanisms using Isogeny Based Algorithms)*

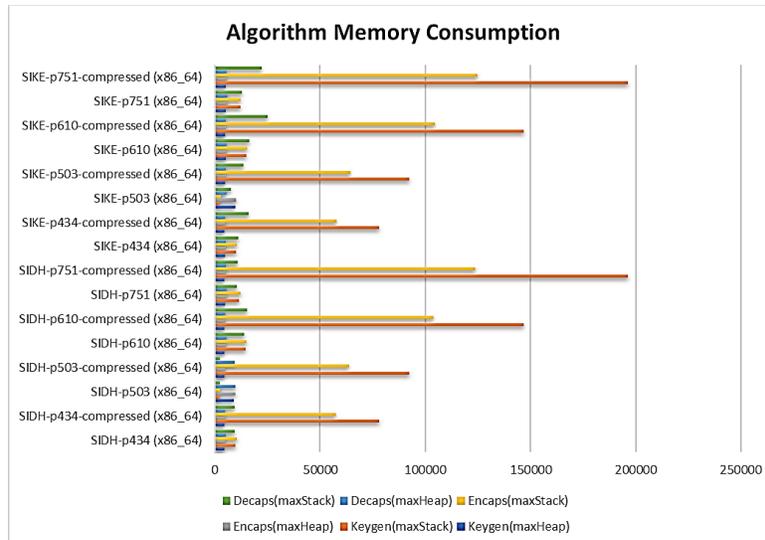

*Figure 21: -Memory consumption (bytes per algorithm)*
*(Key Encapsulation Mechanisms using Isogeny Based Algorithms)*

## *6.5 Lattice-Based Cryptography*

Lattice-based cryptography is based on the computational hardness of lattice problems, the basis of which is the shortest vector problem (SVP). Here, we are given an input, lattice represented by an arbitrary basis, and our goal is to output the shortest nonzero vector in it [43]. Simply put, the goal of the attacker is to find the shortest vector from the origin when given the basis of a lattice. A zero vector doesn't work as an answer. The space requirement of this algorithm is exponential and the best-known algorithm has a running time of $2^{O(n \log n)}$ for an exact solution to SVP or even just an approximation to within poly(n) factors. Lattice-based cryptography provides a promising solution for post-quantum cryptography [38][41]. The algorithm is relatively efficient in implementation and has very strong security proofs based on worst-case hardness. The potential quantum-safe algorithm based on the lattice are as follows: -

i) **CRYSTALS-KYBER** encompasses two cryptographic primitives: Kyber, an IND-CCA2-secure key-encapsulation mechanism (KEM); and Dilithium, a strongly EUF-CMA-secure digital signature algorithm. Both the algorithms are based on the hard problem using lattice. The algorithm is computationally fast and has a small public key [22]. It has different parameters set to match the NIST security levels 1, 3, and 5.

ii) **NTRU** is lattice-based cryptography. It was developed in 1996 by mathematicians Jeffrey Hoffstein, Jill Pipher, and Joseph H. Silverman. In 2013, Damien Stehle and Ron Steinfeld developed a post-quantum secure version of NTRU. It is based on the hardness of solving the Ring-learning-with-errors problem (Ring-LWE). The algorithm is resistant to attacks using Shor's algorithm [58].

iii) **SABER** also belongs to the lattice-based cryptography algorithm family. The security of the algorithm relies on the hardness of the Module Learning with Rounding problem (MLWR). SABER is a simple and efficient algorithm [48]. The SABER-suite offers three security levels:
   - LightSABER (NIST Level 1)
   - SABER (NIST Level 3)
   - FireSABER (NIST Level 5)

iv) **FrodoKEM** is based on the algebraically unstructured lattice. The algorithm has fewer parameter restrictions but a large public key size [29]. FrodoKEM is designed for IND-CCA security at three levels:
- FrodoKEM-640, which targets Level 1 in the NIST
- FrodoKEM-976, which targets Level 3 in the NIST
- FrodoKEM-1344, which targets Level 5 in the NIST

v) **NTRU Prime** is an improvised version of the original NTRU which has eliminated many of the complications of the lattice security review [44].

vi) **CRYSTALS-DILITHIUM i**s based on the "Fiat-Shamir with Aborts" technique of Lyubashevsky which uses rejection sampling to make lattice-based Fiat-Shamir schemes compact and secure [24]. Dilithium claims to have the smallest public key and signature size of any lattice-based signature scheme that only uses uniform sampling.

vii) **FALCON** is a lattice-based signature scheme designed on the theoretical framework of Gentry, Peikert, and Vaikuntanathan. The security of the algorithm is based on the underlying hard problem of the short integer solution problem (SIS) over NTRU lattices, for which no efficient solving algorithm is currently known in the general case, even with the help of quantum computers [26].

Lattice-based cryptography algorithms seem to be the most promising and quantum-safe. The maximum number of algorithms announced by NIST in 3rd round belongs to the lattice-based cryptography family (Table 14). There are three algorithms' CRYSTALS-KYBER, NTRU and SABER declared as a finalist for Key Encapsulation Mechanisms. CRYSTALS-DILITHIUM and FALCON are declared as a finalist for digital signatures. FrodoKEM and NTRU Prime have been announced as alternate candidates for Key Encapsulation Mechanisms. The runtime behavior and memory consumptions of the algorithms are shown in Table 15, Table 16, Table 17, and Table 18.

*Table 14: - Lattice-Based Quantum-Safe Cryptographic Algorithms*

| Sl. No. | Algorithm Name | Current Status | Open-Source C Library for Quantum-Safe Cryptographic Algorithms (liboqs) | | | | Usage |
|---|---|---|---|---|---|---|---|
| | | | NIST Level | Algorithms Name | Public Key Size (bytes) | Private Key Size (bytes) | |
| 1 | CRYSTALS-KYBER | Finalists | 1 | Kyber512 | 800 | 1632 | Key Encapsulation Mechanisms |
| | | | 1 | Kyber512-90s | 800 | 1632 | |
| | | | 3 | Kyber768 | 1184 | 2400 | |
| | | | 3 | Kyber768-90s | 1184 | 2400 | |
| | | | 5 | Kyber1024 | 1568 | 3168 | |
| | | | 5 | Kyber1024-90s | 1568 | 3168 | |
| 2 | NTRU | Finalists | 1 | NTRU-HPS-2048-509 | 699 | 935 | Key Encapsulation Mechanisms |
| | | | 3 | NTRU-HPS-2048-677 | 930 | 1234 | |
| | | | 3 | NTRU-HRSS-701 | 1138 | 1450 | |
| | | | 5 | NTRU-HPS-4096-821 | 1230 | 1590 | |
| 3 | SABER | Finalists | 1 | LightSaber-KEM | 672 | 1568 | Key Encapsulation Mechanisms |
| | | | 3 | Saber-KEM | 992 | 2304 | |
| | | | 5 | FireSaber-KEM | 1312 | 3040 | |
| 4 | FrodoKEM | Alternates | 1 | FrodoKEM-640-AES | 9616 | 19888 | |

| | | | | FrodoKEM-640-SHAKE | 9616 | 19888 | |
|---|---|---|---|---|---|---|---|
| | | | 3 | FrodoKEM-976-AES | 15632 | 31296 | Key Encapsulation Mechanisms |
| | | | | FrodoKEM-976-SHAKE | 15632 | 31296 | |
| | | | 5 | FrodoKEM-1344-AES | 21520 | 43088 | |
| | | | | FrodoKEM-1344-SHAKE | 21520 | 43088 | |
| 5 | NTRU Prime | Alternate | 1 | ntrulpr653 | 897 | 1125 | Key Encapsulation Mechanisms |
| | | | | sntrup653 | 994 | 1518 | |
| | | | 2 | ntrulpr761 | 1039 | 1294 | |
| | | | | sntrup761 | 1158 | 1763 | |
| | | | 3 | ntrulpr857 | 1184 | 1463 | |
| | | | | sntrup857 | 1322 | 1999 | |
| 6 | CRYSTALS-DILITHIUM | Finalists | 2 | Dilithium2 | 1312 | 2528 | Digital Signature Algorithms |
| | | | | Dilithium2-AES | 1312 | 2528 | |
| | | | 3 | Dilithium3 | 1952 | 4000 | |
| | | | | Dilithium3-AES | 1952 | 4000 | |
| | | | 5 | Dilithium5 | 2592 | 4864 | |
| | | | | Dilithium5-AES | 2592 | 4864 | |
| 7 | FALCON | Finalists | 1 | Falcon-512 | 897 | 1281 | Digital Signature Algorithms |
| | | | 5 | Falcon-1024 | 1793 | 2305 | |

***Table 15: -*** *Runtime Analysis of Open Quantum Safe Lattice-Based Cryptographic Algorithms (Key Encapsulation Mechanisms)*

| Algorithm | Keygen/s | Keygen (cycles) | Encaps/s | Encaps (cycles) | Decaps/s | Decaps (cycles) |
|---|---|---|---|---|---|---|
| FireSaber-KEM (x86_64) | 21419.7 | 116577 | 20332.3 | 122807 | 21608 | 115601 |
| FrodoKEM-1344-AES (x86_64) | 589 | 4244465 | 463.18 | 5397844 | 475.17 | 5261594 |
| FrodoKEM-1344-SHAKE (x86_64) | 200.07 | 12494691 | 192.33 | 12999407 | 192.87 | 12962247 |
| FrodoKEM-640-AES (x86_64) | 2179 | 1147030 | 1578 | 1584240 | 1653 | 1512386 |
| FrodoKEM-640-SHAKE (x86_64) | 794 | 3148801 | 669.11 | 3735711 | 719.76 | 3473037 |
| FrodoKEM-976-AES (x86_64) | 1041.32 | 2400233 | 783.74 | 3189368 | 816 | 3063901 |
| FrodoKEM-976-SHAKE (x86_64) | 371.42 | 6731021 | 340.22 | 7348775 | 341.66 | 7316407 |
| Kyber1024 (x86_64) | 35936 | 69446 | 31185.3 | 80016 | 41011.33 | 60859 |
| Kyber1024-90s (x86_64) | 51866.3 | 48061 | 45639 | 54625 | 67317 | 37038 |
| Kyber512 (x86_64) | 67526 | 36882 | 58398.7 | 42674 | 92577.67 | 26910 |
| Kyber512-90s (x86_64) | 86699 | 28696 | 81124.7 | 30692 | 135048.3 | 18406 |
| Kyber768 (x86_64) | 46470.7 | 53657 | 42638.7 | 58493 | 59629.33 | 41825 |
| Kyber768-90s (x86_64) | 66675 | 37353 | 60583.7 | 41115 | 96815.67 | 25724 |
| LightSaber-KEM (x86_64) | 44989.7 | 55426 | 46295 | 53860 | 53519.67 | 46617 |
| NTRU-HPS-2048-509 (x86_64) | 14699 | 169936 | 60072.3 | 41464 | 71687.33 | 34780 |
| NTRU-HPS-2048-677 (x86_64) | 8829 | 283020 | 46316.7 | 53816 | 48695.67 | 51245 |

| Algorithm | | | | | | |
|---|---|---|---|---|---|---|
| NTRU-HPS-4096-821 (x86_64) | 6555.67 | 381217 | 39904.7 | 62494 | 38197.33 | 65355 |
| NTRU-HRSS-701 (x86_64) | 9552.67 | 261573 | 66261.7 | 37568 | 46262.67 | 53946 |
| Saber-KEM (x86_64) | 29017.7 | 86015 | 29338.3 | 85067 | 32348 | 77189 |
| ntrulpr653 (x86_64) | 32654.3 | 76422 | 29245 | 85325 | 27985.67 | 89232 |
| ntrulpr761 (x86_64) | 31496.3 | 79234 | 30106.7 | 82886 | 27409.33 | 91111 |
| ntrulpr857 (x86_64) | 26871 | 92888 | 24131.7 | 103440 | 20983.33 | 119045 |
| sntrup653 (x86_64) | 2980.33 | 838768 | 37728 | 66108 | 37468.33 | 66618 |
| sntrup761 (x86_64) | 2349.33 | 1064025 | 34414.7 | 72484 | 37841 | 65973 |
| sntrup857 (x86_64) | 1814.33 | 1377979 | 29393.7 | 84894 | 28335.67 | 88127 |

*Table 16: -* *Memory Use Analysis (bytes per algorithm) of Open Quantum Safe Lattice-Based Cryptographic Algorithms (Key Encapsulation Mechanisms)*

| Algorithm | Keygen (maxHeap) | Keygen (maxStack) | Encaps (maxHeap) | Encaps (maxStack) | Decaps (maxHeap) | Decaps (maxStack) |
|---|---|---|---|---|---|---|
| FireSaber-KEM (x86_64) | 8752 | 25744 | 10256 | 25904 | 10288 | 27408 |
| FrodoKEM-1344-AES (x86_64) | 68832 | 88512 | 90448 | 173352 | 90480 | 216616 |
| FrodoKEM-1344-SHAKE (x86_64) | 69616 | 77352 | 91280 | 123720 | 90480 | 165704 |
| FrodoKEM-640-AES (x86_64) | 34160 | 43392 | 43464 | 84008 | 43480 | 104616 |
| FrodoKEM-640-SHAKE (x86_64) | 34512 | 37896 | 44248 | 58624 | 44264 | 79264 |
| FrodoKEM-976-AES (x86_64) | 51152 | 64616 | 67088 | 128064 | 67376 | 159488 |
| FrodoKEM-976-SHAKE (x86_64) | 51936 | 55712 | 67704 | 89568 | 67728 | 121024 |
| Kyber1024 (x86_64) | 13560 | 16720 | 11344 | 21944 | 11376 | 23544 |
| Kyber1024-90s (x86_64) | 13560 | 17168 | 10512 | 19928 | 10544 | 21528 |
| Kyber512 (x86_64) | 11256 | 7504 | 8240 | 11704 | 8272 | 12504 |
| Kyber512-90s (x86_64) | 11256 | 7952 | 7408 | 9720 | 7440 | 10520 |
| Kyber768 (x86_64) | 12408 | 11600 | 9712 | 16344 | 9744 | 17464 |
| Kyber768-90s (x86_64) | 12408 | 11984 | 8880 | 14264 | 8912 | 15384 |
| LightSaber-KEM (x86_64) | 6416 | 14776 | 7184 | 14936 | 7216 | 15704 |
| NTRU-HPS-2048-509 (x86_64) | 5810 | 35744 | 6541 | 30816 | 6573 | 29120 |
| NTRU-HPS-2048-677 (x86_64) | 6340 | 50944 | 7302 | 44160 | 7334 | 41888 |
| NTRU-HPS-4096-821 (x86_64) | 6996 | 64320 | 8258 | 56288 | 8290 | 53632 |
| NTRU-HRSS-701 (x86_64) | 6764 | 49136 | 7934 | 42352 | 7966 | 42128 |
| Saber-KEM (x86_64) | 7472 | 19928 | 8592 | 20088 | 8624 | 21208 |
| ntrulpr653 (x86_64) | 6198 | 26328 | 7255 | 26552 | 7287 | 27864 |
| ntrulpr761 (x86_64) | 6509 | 27512 | 7708 | 27672 | 7740 | 29208 |
| ntrulpr857 (x86_64) | 6823 | 34968 | 8167 | 35064 | 8199 | 36760 |

| Algorithm | | | | | | |
|---|---|---|---|---|---|---|
| sntrup653 (x86_64) | 6688 | 23064 | 7617 | 24120 | 7649 | 24152 |
| sntrup761 (x86_64) | 7097 | 23672 | 8168 | 24888 | 8200 | 24984 |
| sntrup857 (x86_64) | 7497 | 30616 | 8713 | 31960 | 8745 | 32056 |

*Table 17: - Runtime Analysis of Open Quantum Safe Lattice-Based Cryptographic Algorithms (Digital Signature Algorithms)*

| Algorithm | Keypair/s | Keypair (cycles) | Sign/s | Sign (cycles) | Verify/s | Verify (cycles) |
|---|---|---|---|---|---|---|
| Dilithium2 (x86_64) | 26417.7 | 94494 | 10520 | 237479 | 29132 | 85721 |
| Dilithium2-AES (x86_64) | 44800 | 55671 | 14407.7 | 173353 | 44537.33 | 56039 |
| Dilithium3 (x86_64) | 16110.3 | 155032 | 6506.33 | 384098 | 17561.33 | 142256 |
| Dilithium3-AES (x86_64) | 29631 | 84225 | 9637.67 | 259239 | 30018.33 | 83182 |
| Dilithium5 (x86_64) | 10141.7 | 246360 | 5279.67 | 473377 | 10341 | 241650 |
| Dilithium5-AES (x86_64) | 19582 | 127496 | 8141.33 | 306924 | 19511.33 | 128032 |
| Falcon-1024 (x86_64) | 40.23 | 62157513 | 1446.52 | 1727907 | 9782.67 | 255449 |
| Falcon-512 (x86_64) | 105.05 | 23797285 | 2890.33 | 864862 | 19684.33 | 126886 |

*Table 18: - Memory Use Analysis (bytes per algorithm) of Open Quantum Safe Lattice-Based Cryptographic Algorithms (Digital Signature Algorithms)*

| Algorithm | Keygen (maxHeap) | Keygen (maxStack) | Sign (maxHeap) | Sign (maxStack) | Verify (maxHeap) | Verify (maxStack) |
|---|---|---|---|---|---|---|
| Dilithium2 (x86_64) | 12656 | 20880 | 11584 | 49616 | 11360 | 21592 |
| Dilithium2-AES (x86_64) | 12656 | 17040 | 15400 | 44816 | 10528 | 14800 |
| Dilithium3 (x86_64) | 14768 | 25656 | 14569 | 78808 | 14345 | 24728 |
| Dilithium3-AES (x86_64) | 14768 | 20792 | 18385 | 69424 | 13513 | 16848 |
| Dilithium5 (x86_64) | 16272 | 33848 | 17375 | 113232 | 17151 | 31128 |
| Dilithium5-AES (x86_64) | 16272 | 26936 | 21191 | 108336 | 16319 | 21264 |
| Falcon-1024 (x86_64) | 12914 | 36792 | 14344 | 80880 | 9696 | 10512 |
| Falcon-512 (x86_64) | 10994 | 19216 | 11784 | 40944 | 11784 | 2320 |

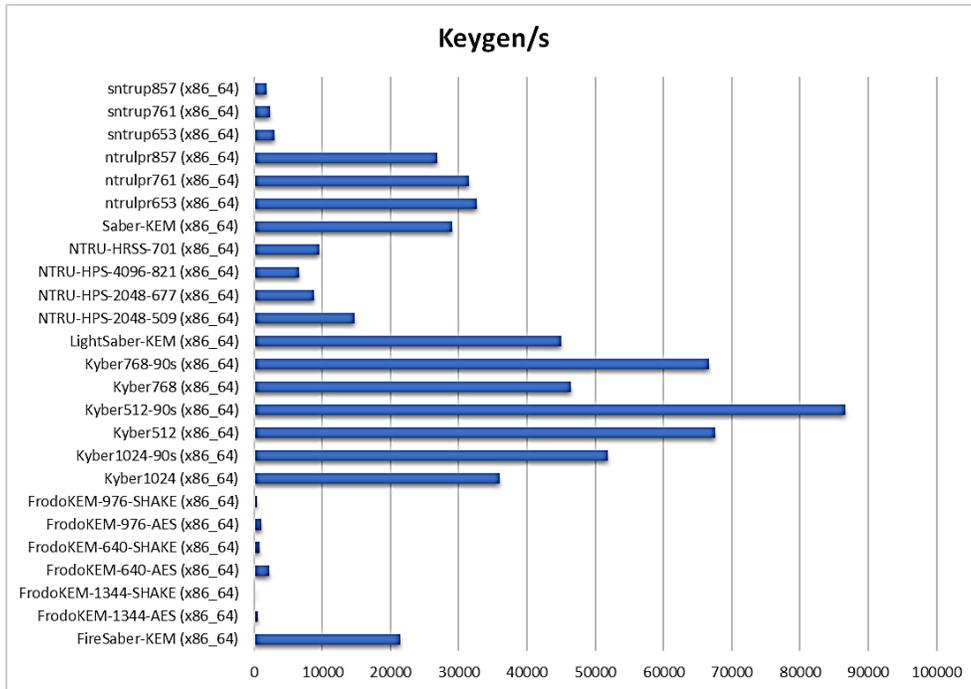

*Figure 22: - Keygen operations per second per algorithm (Key Encapsulation Mechanisms using Lattice Based Algorithms)*

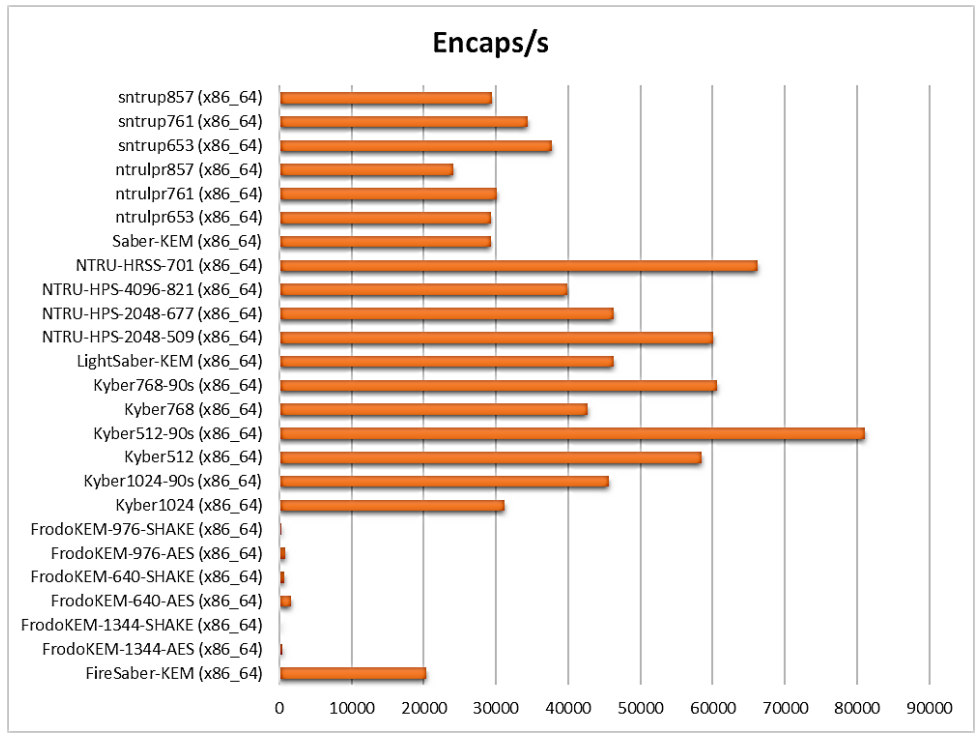

*Figure 23: - Encaps operations per second per algorithm (Key Encapsulation Mechanisms using Lattice Based Algorithms)*

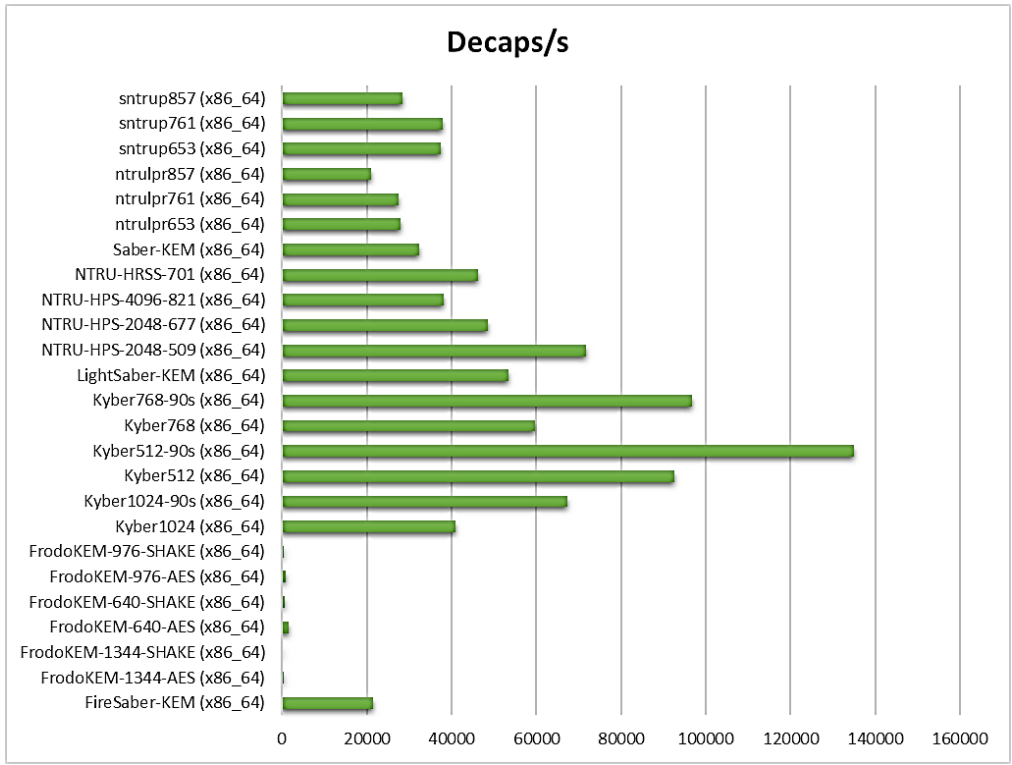

*Figure 24: - Decaps operations per second per algorithm (Key Encapsulation Mechanisms using Lattice Based Algorithms)*

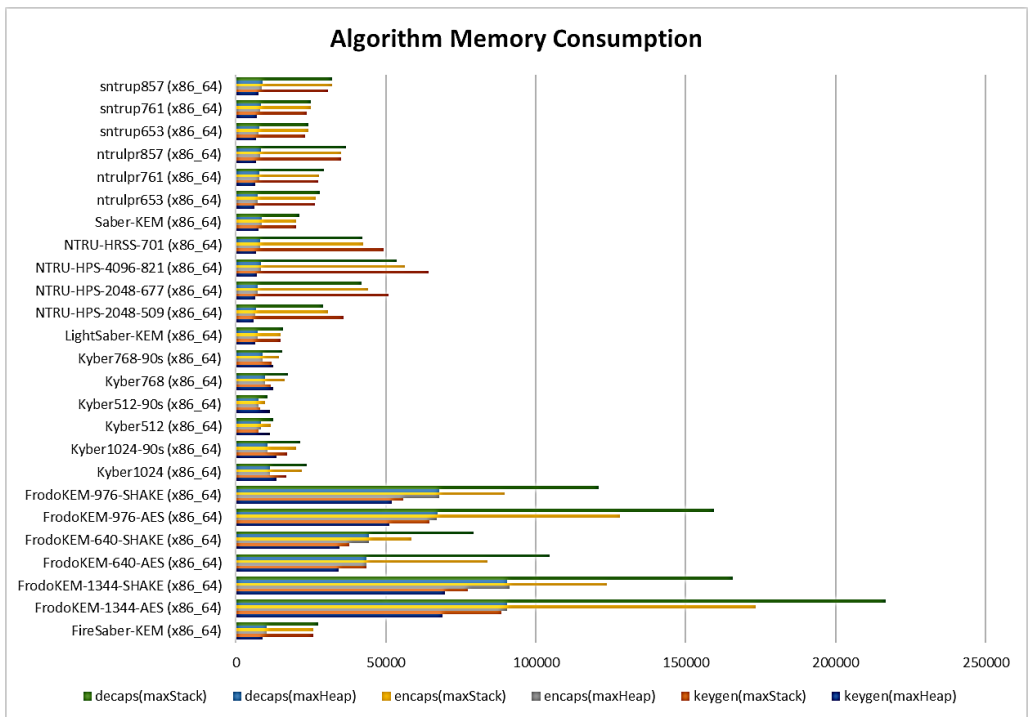

*Figure 25: - Memory consumption (bytes per algorithm) (Key Encapsulation Mechanisms using Lattice Based Algorithms)*

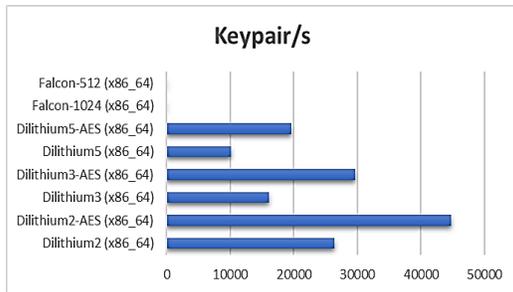

*Figure 26: -Keypair operations per second per algorithm (Digital Signature using Lattice Based Algorithms)*

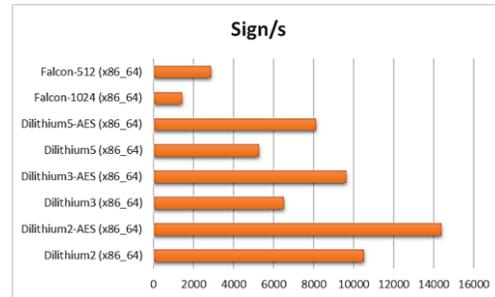

*Figure 27: - Sign operations per second per algorithm (Digital Signature using Lattice Based Algorithms)*

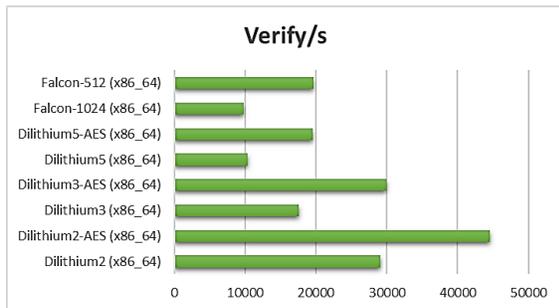

*Figure 28: - Verify operations per second per algorithm (Digital Signature using Lattice Based Algorithms)*

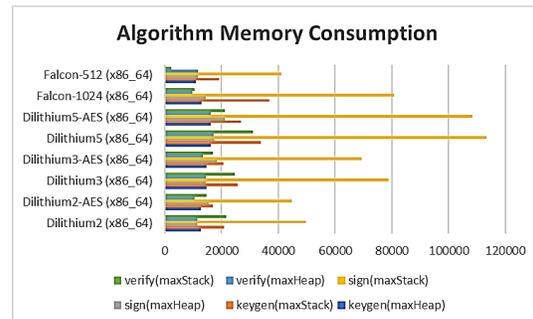

*Figure 29: - Memory consumption (bytes per algorithm) (Digital Signature using Lattice Based Algorithms)*

## 7. Implementation Feasibility and Migration Challenges

Performance analysis of various quantum-safe algorithms shows that in general, quantum-safe algorithms require many CPU cycles for basic operations (i.e., key generation, encrypt/decrypt, key exchange, sign, verify, etc.). The algorithms also require large to the very large public key and private key sizes. The runtime memory consumption of these algorithms is very high compared to the classical cryptographic algorithms.

The higher CPU cycle and memory usage is constrained to the ICT systems and devices. The resource constraints can be resolved for the systems like a laptop, desktop, or high-end server up to some extent. However, it's a challenging issue for small devices like smartphones, sensor networks, smart-grid, IoT, smart devices, smart homes, etc.

Security is paramount for all types of ICT devices. Hence the implementation feasibility of quantum-safe algorithms is very important. The performance analysis of the post-quantum cryptography algorithms announced by NIST 3rd round shows that only a few algorithms i.e., SIKE, NTRU, and NTRU Prime could be used for the Key Encapsulation Mechanisms and SPHINCS+ could be used for the digital signature purpose on the small devices.

Adoption and implementation of these algorithms will depend on various factors such as throughput, latency threshold, key size limitations of current hardware and software, etc. The replacement of classical cryptography algorithms with quantum-safe algorithms will require changes in existing protocol, communication devices, hardware with high processing capacity to accelerate the algorithm performance, application code, operating system, etc.

NIST is still in the process of finalizing Quantum-Safe algorithms. Most likely the process will continue to the fourth round of evaluation. The algorithms which are declared as a finalist has acceptable performance but require additional analysis to gain sufficient confidence. Some of the algorithms which are declared as alternate may potentially be standardized based on the high confidence in their security.

Currently, the prime objective is to develop a quantum-resistant algorithm to protect from a possible future attack. However, major implementation feasibility issues which require intense effort to overcome are: -

- **Migration Overhead and Complexity: -** Migration to PQC will require significant upgradation in any organization's current cryptographic infrastructure. Some of their current IT components may become completely obsolete and will require replacement. The existing algorithms, software, and protocol may require redesign and changes. Overall, the entire migration process will add substantial budget overhead to the organization and unavoidable complexity.
- **Scalability Issue: -** Scalability is a challenging issue for most of the potential PQC algorithms. It is difficult to prove the hardness of the algorithm at scale. For example, lattice-based cryptography which is one of the popular techniques used for designing quantum-safe algorithms does manage to scale well but achieve average-case hardness. There is a trade-off between hardness and scalability. Either can be achieved, but not both.
- **Large Key Size and Speed: -** Most of the potential PQC algorithms require large to very large key sizes compared to the existing cryptography algorithms. Because of the large key size and complex algorithm, more time is required to encrypt, decrypt or verify the message at either end. It also requires more storage space in the device and a longer time to establish secure communication.
- **Resource-Constrained Devices: -** Smartphones, IoT Devices, Sensor Networks, Industrial Control Systems, and SCADA are some of the most critical components of our infrastructure. In general, these devices have less memory and processing capacity that poses a challenge for strong security implementation. Post-quantum cryptography algorithms require intense customization and feasibility study for their implementation in the resource-constrained device. It's tough to maintain the hardness of the cryptography algorithm while satisfying the resource constraints.

## 8. Conclusions

To break the current cryptographic algorithms, a quantum computer with millions of 'Qubits' is required. It's not sure that when the quantum computers with a scale of a million qubit will be available in the commonplace. However, the exponential growth in quantum computer technology's development shows that the storm is approaching very fast. The information we communicate today is mostly encrypted using a public-key cryptography algorithm that is vulnerable to quantum computers. It's quite possible that many hackers, state-sponsored or nation-states themselves might be intercepting and harvesting the encrypted messages with envisioning that they will be able to decrypt these messages when the quantum computing resources will be available in the future. To counter this threat, we need to migrate to post-quantum cryptographic algorithms at the earliest. Migration to post-quantum cryptographic algorithms requires thorough planning and implementation globally. The migration process will require work on the various front including education, training, and awareness. It will require extensive coordination among many stakeholders involved in the design and development of hardware, software, and IT infrastructure components. Early preparation will ensure a less costly and effective changeover with minimal disruption.

# DECLARATIONS

**Funding: -** Not applicable

**Conflicts of interest/Competing interests**: - The author hereby declares that they have no conflict of interest. No research grant or fund has been received from any agency to carry out the research work discussed in the manuscript.

**Availability of data and material** (data transparency): - Data sharing is not applicable to this article as no new data were created or analyzed in this study.

**Code availability**: - Not applicable

**Authors' contributions**: - The paper is authored by a single author and all the works in the paper are carried out by him.

**Ethics approval**: - Not applicable

**Consent to participate**: - Not applicable

**Consent for publication**: - Not applicable

# COMPLIANCE WITH ETHICAL STANDARDS

This article does not contain any studies with human participants or animals performed by any of the authors.